\documentclass[draft]{agujournal2019}
\usepackage{url} %this package should fix any errors with URLs in refs.
\usepackage[inline]{trackchanges} %for better track changes. finalnew option will compile document with changes incorporated.
\usepackage{soul}
\usepackage{amssymb}
\nolinenumbers    %need this to comply with Arxiv req.
\newcommand{\bvec}[1]{\mathbf{#1}}

\newcommand{\comment}[1]{}
\newcommand{\red}[1]{#1}

\draftfalse
\journalname{JGR: Space Physics}
\begin{document}

\title{Observations of Energized Electrons in the Martian Magnetosheath}

\authors{K.~Horaites\affil{1}, L.~Andersson\affil{1}, S.~J.~Schwartz\affil{1}, S.~Xu\affil{2}, D.~L.~Mitchell\affil{2}, C.~Mazelle\affil{3}, J.~Halekas\affil{4}, J.~Gruesbeck\affil{5}}

 \affiliation{1}{Laboratory for Atmospheric and Space Physics, Boulder, CO, USA}
 \affiliation{2}{Space Sciences Laboratory, Berkeley, CA, USA}
 \affiliation{3}{IRAP CNRS-University of Toulouse-UPS-CNES, Toulouse, France}
 \affiliation{4}{Department of Physics and Astronomy, University of Iowa, Iowa City, IA, USA}
 \affiliation{5}{Goddard Space Flight Center, Greenbelt, Maryland, USA}

\begin{keypoints}
% max 140 characters

\item The energization of electrons crossing the Martian bow shock can be described in terms of a quasi-static localized potential drop.

\item The energy difference of electrons entering the Martian bow shock at opposite locations on the same flux tube is much less than expected.

\item A global, distributed potential within the sheath could resolve the unexpected trends reported in this work.

\end{keypoints}

%% -------------------------------------------------------------- %%

%% \begin{abstract} starts the second page

\begin{abstract}

%The energization of electrons in the Martian magnetosheath is investigated in this paper. The magnitude and location of the energization of sheath electrons has been proposed in other studies, and 
\red{This observational study demonstrates that the magnitude and location of energization of electrons in the Martian magnetosheath is more complex than previous studies suggest.}  
Electrons in Mars's magnetosheath originate in the solar wind and are accelerated by an electric field when they cross the bow shock.
%\red{Previous studies have assumed that the magnitude of this acceleration can be explained by classic bow shock theory.} 
Assuming that this acceleration is localized solely to the shock, the field-aligned electron distributions in the sheath are expected to be highly asymmetric. However, such an asymmetry is not observed in this study. Based on the analysis here, it is suggested that an additional parallel acceleration takes place downstream of the Martian bow shock. This additional acceleration suppresses the expected asymmetry of the electron distribution.
 Consequently, along a flux tube in the magnetosheath that is tied on both ends to the bow shock the difference in energization between parallel and anti-parallel electrons is less than about 20~eV.
Where this energization difference is expected to be maximal, we find the energization difference to be at most $\lesssim$25\% of the predicted value. 
 %. We propose this is driven by either current limitation or a largescale potential structure. Of the two, the latter is suggested to be more likely set up by the electron pressure inside the bow shock.
\end{abstract}

\section*{Plain Language Summary}

As the supersonic solar wind plasma encounters an obstacle, it is first slowed down to subsonic speeds and then diverted around the object. 
%At the bow shock, i.e. the surface where the flow becomes subsonic, the ions are slowed down while the electrons are accelerated to higher speeds.
%At the bow shock, i.e. the surface where the flow becomes subsonic, individual electrons are accelerated by and electric field within the shock.   %sjs
At the shock wave ahead of a planet, called the planet's ``bow shock'', individual electrons are accelerated by and electric field within the shock.   %sjs
 These energized electrons move quickly along the local magnetic field from one side of the bow shock to the other. Downstream of the bow shock, the two electron populations moving in opposite directions along the magnetic field line should then have crossed the bow shock at the two locations where the field line meets the shock. Since the amount of energy gained by electrons is in general different at the two crossing locations, the two streaming electron populations observed downstream of the bow shock are expected to be energized by different amounts. 
%On the contrary, this study identifies that the two populations are energized very similarly, suggesting that additional acceleration downstream of the` bow shock is required. This paper suggests two viable mechanisms that could explain the observations.
On the contrary, this study identifies that away from the shock the two populations appear to have been energized very similarly. This may imply an additional acceleration downstream of the bow shock is required. This paper suggests two viable mechanisms that could explain the observations.

\section{Introduction}\label{intro_sec}

When the solar wind encounters an obstacle, the bulk flow is decelerated at the bow shock to become subsonic. The bulk flow is then decelerated further and diverted behind the bow shock so that the magnetic field lines---which are approximately frozen in to the fluid---``drape'' around the obstacle.  This effect is observed at comets \cite{koenders16} and downstream of planetary bow shocks such as at Earth \cite{spreiter66} and Mars \cite{nagy04, mazelle04}.

The bow shock is the location where the solar wind goes from supersonic to subsonic.
%As the solar wind slows down, i.e. the ion speeds decrease, the electrons instead are accelerated to higher speeds.
As the solar wind slows down, i.e. the \red{bulk speed decreases}, individual electrons are accelerated to higher energies by the cross-shock potential.
 The degree of electron energization is dependent on position at the shock surface: the electron kinetic temperature at the subsolar point in the sheath may reach $\sim$100 eV, while the energization is much less pronounced at the flanks of the shock. This may be compared to the ambient solar wind electron temperature $\sim$10 eV at Mars. 

Earth studies of collisionless shocks (such as bow shocks) are applicable to Mars since the physics of shocks is universal. In a planetary bow shock, electrons are energized in a very thin region ($\lesssim$1 km wide) near the shock, and the electron distributions have a ``flat-top'' shape \cite{montgomery70}. A kinetic model was developed in \citeA{mitchell14} to predict the form of this relatively isotropic feature in Earth's magnetosheath, by propagating electrons along field lines \red{that employed} Rankine-Hugoniot jump conditions, described in e.g.~\citeA{kivelson95}.  The distributions are not perfectly isotropic however, \red{as noted e.g. in \citeA{mitchell12}.} Comparison of the field-parallel and perpendicular temperatures has been used to suggest that anisotropic heating might also take place \cite{feldman83b}. The electron energization is generally believed to be caused by an ambipolar potential \cite{scudder86b, lefebvre07}, which owes to the strong electron pressure gradient across the shock. However, other explanations of electron energization at shocks, such as via turbulent dissipation \cite{sagdeev66, galeev76} have been developed.
%In \cite{lefebvre07}, the electron distributions were studied throughout three bow shock crossings observed by the Cluster spacecraft; the authors found that the spatially-dependent inflation of the electron distributions was consistent with energization provided by and electrostatic ambipolar potential.
 %describing the shock as providing a parallel monotonic energization though out the shock under the assumption of conservation of parallel energy and magnetic moment \cite{schwartz98}.

The region inside the bow shock, known as the sheath, is where the shocked solar wind diverts around the object and further deceleration of the advecting flux tubes takes place. 
At the lowest altitudes in the sheath, a transition region separates the external decelerating solar wind ions from an internal region where plasma processes are controlled by the planetary plasma environment. 
%The location where the solar wind ions are completely diverted around the object is the transition region where inside the plasma processes are controlled by the planet plasma environment. 
The transition from the sheath to the planetary plasma occurs over a region identifiable by multiple observational signatures. 
In this transition region, one can find for example the ``magnetic pileup boundary'' or ``MPB'' \cite{acuna98}, Ion Composition Boundary (e.g., \citeA{halekas19}), and Induced Magnetospheric Boundary \cite{lundin04}. The exact location of the boundary is not crucial for the outcome of this paper, so we will adopt the empirical position of the MPB reported in \citeA{vignes00} to locate this transition region. 
%At Mars this region is referred to by various names, of which one is the ``magnetic pileup boundary'' (MPB). \red{The MPB is roughly co-located with the Induced Magnetospheric Boundary \cite{lundin04} and the Ion Composition Boundary (e.g., \citeA{halekas19}), and each of these boundaries comes with its own criterion for separating the ionosphere from the solar wind-influenced portion of the magnetosphere.
%These distinctions, however, will not be crucial for the present work.}
%The existence of multiple names for the same feature is a symptom of the fact that this transition is not a well-defined boundary as can be found at Earth. Although the MPB lacks a clear definition, this boundary's identification will not be crucial for the present work. 

% KH: This paragraph was moved earlier
In the Martian sheath near the transition region that includes the MPB, the sheath electron distributions were found to be ``eroded'' \cite{crider00}; i.e., the phase space density of energetic electrons (at a given energy $\sim$100eV) sharply decreased over this region, by up to 2 orders of magnitude as compared to higher altitudes in the sheath. 
In that study the erosion was explained by the presence of the neutral Martian corona, which reaches well into the Martian sheath. 
It was suggested that sheath electrons collide with the neutral gas, and the resulting process of electron impact ionization \red{(a process that has also been reported independently in the Martian foreshock, e.g.~\citeA{mazelle18})} causes the electrons to lose energy. This suggestion was critiqued in \citeA{schwartz19}, where it was argued that sheath electrons spend too little time at the highest neutral densities for this process to be of importance. Therefore a collisionless kinetic model was developed describing electrons flowing along a solar wind magnetic flux tube as it drapes around Mars.
% And as the ends of flux tube see different potential acceleration at two different bow shock locations energies of the electron cross talk should be different. 
The model accounted for the non-uniformity of the flux tube deceleration, and also distinguished between electrons that pass through the system and those that are temporarily trapped inside the bow shock. The different electron histories were evaluated, which resulted in eroded distributions that compared favorably with electron distributions observed by the MAVEN spacecraft. We may infer from this recent work that \red{to a first approximation} electrons evolve collisionlessly in the Martian sheath.
%Only one orbit was evaluated however. We also note that the orbit was selected (to agree with the model) during conditions where the solar wind's interaction with Mars would set up a symmetric system; in this special case the electron distributions would be symmetric despite cross-talk. This motivates the need for a more comprehensive study to evaluate the sheath electron energies, and specifically to see if cross talk leads to the development of anisotropies in the electron distributions as expected in the general case.

Such collisionless evolution of electrons has been investigated in the context of Earth's magnetosheath, in \citeA{mitchell12, mitchell13, mitchell14}. These studies emphasized the non-locality of electron kinetics in the sheath. By non-locality we mean the following: since guiding centers of moving electrons are expected to propagate along the magnetic field lines (which in turn advect with the bulk flow), and moreover because the electrons are transported collisionlessly at speeds much greater than the bulk flow speed, the distribution function $f(\bvec{v})$ observed at a given point in the sheath will in general be a convolution of electrons that crossed the bow shock at different locations.  %[This paragraph needs to get a punch line is it the next paragraph that should be merged with the next paragraph?]
 This communication between distant bow shock locations was termed ``electron cross talk''. 

In \citeA{mitchell12}, using Cluster and THEMIS B data it was shown that the electron distributions can exhibit appreciable field-parallel anisotropy. The authors argued that this asymmetry arises because field-parallel and anti-parallel electrons cross the shock at two different locations along the field line, with different cross-shock potentials. Because the magnetic fields at Mars are similarly draped and thread the local bow shock, we may expect cross talk to generate Martian field-parallel electron distributions with this same systematic anisotropy.

%Evaluation of sheath electrons distributions as the magnetic field drapes over Mars is possible because of the small size of the system. This effect is more difficult to study Earth because the large size of the magnetosphere. 
This study investigates if the sheath electrons \red{at Mars carry} information from the bow shock via ``cross talk''. The mission and the data set from MAVEN's SWEA electrostatic analyzer is first presented in section~\ref{swea_sec}.  A statistical study of the energization of the electrons in the sheath is presented in section~\ref{obs_sec}. This study will show the asymmetry of the electron distribution that may be caused by cross talk to be smaller than expected. Processes that could cause these more symmetric electron distributions are suggested in section~\ref{discussion_sec}. The paper is summarized in section \ref{summary_sec}. A detailed presentation of the distribution mapping (used to infer the energization) and error evaluation are provided in the supplementary material to this paper.

\section{Theory}\label{theory_sec}

In a collisionless plasma, the evolution of the distribution function $f(\bvec{v}, \bvec{x}, t)$ obeys {\it Liouville's theorem}.
% which describes how a particle population will develop if only electromagnetic forces are preset (see appendix). 
If the electric and magnetic fields are known along a particle path, one can perform a ``Liouville mapping'' \cite{schwartz98} to predict how the distribution will vary with position along the path.  Conversely, if the particle distributions and magnetic fields at various points along an expected particle path are measured, the electric field along the path can be estimated. Since the variation of magnetic field strength does not influence the pitch angles of particles whose velocities are exactly field-aligned, the field-aligned cuts of the electron distribution are only influenced by the electric field. For particles with a significant perpendicular velocity component, the magnetic field gradients should also be considered when performing a Liouville mapping. This methodology is commonly applied assuming the conservation of magnetic moment, steady-state fields and particle distributions, and the absence of collisions, as in e.g. \citeA{lefebvre07}.  

In the process of Liouville mapping, one must take care to distinguish between the ``passing'' and ``reflected'' populations. Both electric fields and magnetic field gradients can reflect particles, which may lead the distributions to develop a loss cone. The term ``loss cone'' usually refers to particles of certain pitch angles but also there are also regions in energy which are excluded. Therefore, when implementing Liouville mapping only the part of the distributions that can be observed at the two locations should be considered; only the portions outside the excluded in pitch angles and energies should be evaluated. 

%The bow shock decelerates the solar wind (i.e. the ions), and on large scales this behavior is due to the cross-shock potential.
\red{The sheath is populated by energized solar wind electrons.
When they cross the bow shock, these electrons receive a net acceleration that can be attributed to the frame-invariant ambipolar component of the cross-shock potential. }
The size of the potential depends primarily on the solar wind conditions and the angle of the solar wind flow vector relative to the shock normal. %\citation{[School book REF].} 
%The solar wind electrons crossing the bow shock are therefore accelerated by this potential and the sheath is populated by energized solar wind electrons.
 The low-energy region of \red{the electron distribution} typically exhibits a ``flat top'' ($f=$const.) shape in Mars's magnetosheath \cite{crider00}; electron distributions in Earth's magnetosheath exhibit a similar feature \cite{feldman83a}.  The energy at which the flat-top ``breaks'' may be used to roughly estimate the degree of energization.

%[I think this should be kept and be a separate paragraph when Introducing Figure 4d i.e. move down%]
%In this study the behavior of the bow shock potential follows the simplistic description presented in \cite{schwartz19}. In that work the solar wind magnetic field is assumed to be straight with a constant cone angle. In the present study we will consider the case where the cone angle is given by the typical Parker angle at Mars ($\sim$60$^\circ$), which is expected to lead to significant asymmetry in the electron distribution due to cross talk. 
%% to the subsolar point of Mars. 
%The shape of the bow shock itself is assumed to follow the empirically derived equation of \cite{vignes00}, but expressed in a simplified parabolic form as in \cite{schwartz19}.

%Inside the bow shock, the magnetic field curves and no longer advects at the solar wind speed. 
Due to their high speeds the electrons will approximately follow trajectories along the instantaneous draped magnetic field. 
%as described in \cite{schwartz19}. 
A kinetic theory of electrons in the Martian magnetosheath was developed in \citeA{schwartz19}; we note that in that study the cone angle $\theta_c$ of magnetic field was assumed to be exactly $90^\circ$. The cone angle is defined here as follows:

\begin{equation}\label{theta_c_eq}
\theta_c \equiv \cos^{-1}(\bvec{B} \cdot \bvec{v}_{sw} / B v_{sw}),
\end{equation}

\noindent where $\bvec{B}$ is the upstream magnetic field and $\bvec{v}_{sw}$ is the solar wind velocity. \red{Note from the definition \ref{theta_c_eq}, we have 0$^\circ$$<$$\theta_c$$<$180$^\circ$; the range $\theta_c$$<$90$^\circ$ corresponds with an anti-sunward pointing upstream field.} Assuming a cylindrically symmetric bow shock, when the cone angle is not exactly 90$^\circ$, the sheath electrons on the same field line originating from two different ends will in general have experienced different cross-shock potentials \red{(see supplementary document for details)}.

In this study the observed field-parallel anisotropy of the electron distribution function will be parametrized by the quantity $\Delta \Phi$:

\begin{equation}\label{delta_phi_eq}
\Delta \Phi = \Phi_\parallel - \Phi_\downarrow.
\end{equation}

\noindent In Eq.~(\ref{delta_phi_eq}) the quantities $\Phi_\parallel$ and $\Phi_\downarrow$ respectively denote the apparent constant energization of the parallel and anti-parallel propagating electron populations. As described above, if the electron energization occurs solely at the bow shock, we should expect the difference $\Delta \Phi$ to be non-zero in general. This motivates the present study, where we will investigate statistically if electrons in the sheath retain information of where they crossed the bow shock.

\section{MAVEN SWEA Electrostatic Analyzer}\label{swea_sec}

The MAVEN (Mars Atmosphere and Volatile Evolution) mission's primary focus is to study the Martian atmosphere \cite{jakosky15}. As a result the satellite includes a comprehensive suite of instruments capable of providing high-quality measurements of the space plasma environment near Mars. In this paper, we will focus on measurements of the electron velocity distribution  provided by MAVEN's Solar Wind Electron Analyzer (SWEA) instrument \cite{mitchell16}. The SWEA instrument \red{has} an energy resolution $\Delta \mathcal{E} / \mathcal{E}$ (FWHM) = 17\%, \red{providing} 79\% coverage of the sky at $\sim$$7^\circ$$\times$22.5$^\circ$ angular resolution.
 Magnetic field observations are made by the MAG magnetometer \cite{connerney15}. 
%To get the solar wind speed the moment information from the Solar Wind Ion Analyzer (SWIA) onboard MAVEN \cite{halekas15} is used. 
The moment information from the Solar Wind Ion Analyzer (SWIA) onboard MAVEN \cite{halekas15} is used to get the solar wind speed.  

In this study, we will consider pitch angle distributions (PADs) computed onboard the MAVEN satellite. These ``survey'' data were regularly sampled by the SWEA instrument, with a time cadence of $\sim$2 seconds with 32 distinct energy steps. \red{These energies are given in the spacecraft frame, which for the fast-moving \red{($\gtrsim$)30 eV} electrons considered here is nearly identical to the Mars rest frame---the frame assumed in our calculations, see supplementary document for details.} At each energy, an automated algorithm chooses 16 different angular positions in phase space (azimuth+elevation pairs) that were sampled by the detector during the accumulation period; these angular positions are so chosen as to lie roughly on a great circle that intersects the local instantaneous magnetic field direction determined by the MAG instrument. The pitch angle is calculated onboard, by comparison with the contemporaneously measured magnetic field provided by MAG.
%SWEA was mounted to optimally capture the field aligned electrons in the solar wind.
For the study here most of the time at least one sector was within $\sim$15$^\circ$ of the magnetic field.

The data considered here cover the time range January 1, 2015 to May 15, 2019. The MAVEN spacecraft orbits Mars in an inclined ellipse with a nominal periapsis altitude targeting a pressure corridor at 150-180 km and an apoapsis altitude of 6220 km \cite{jakosky15}, resulting in an orbit period of 4.5 hours. The subsolar point of the bow shock is located approximately at 2200 km, well within MAVEN's orbit. However, over the Martian year the apoapsis moves from being in front of the planet in the solar wind to deep into the tail of the planet. Consequently there are time periods where MAVEN never crosses the bow shock into the solar wind. For this study only orbits where the satellite reaches well into the solar wind are included.

\section{Observations}\label{obs_sec}

\begin{figure}
\includegraphics[width=1\linewidth]{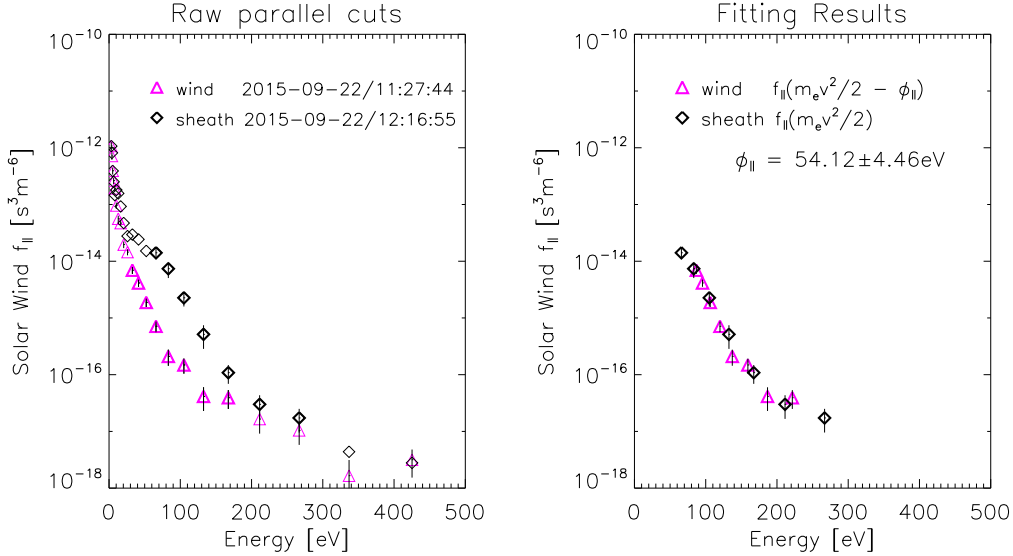}
\caption{\label{fit_example_fig} An example of the parallel cuts in the solar wind and sheath, showing the results of the Liouville mapping procedure. Left: field-parallel cuts $f_\parallel$ of the electron distribution, plotted at an example time in the sheath (black diamonds) and a nearby time in the solar wind (purple triangles) on September 22, 2015. The symbols in bold represent the data that satisfy our selection criteria; these data are used to interpolate the two spectra and calculate their energy difference $\Delta\Phi_\parallel$ (see supplementary document for details). Right: The raw magnetosheath spectrum is again plotted (black diamonds, bold), as well as the same solar wind distribution shifted by the fitted energization $\Phi_\parallel$=54$\pm$4 eV (purple triangles, bold). We observe that when the solar wind spectrum is shifted by this energy $\Phi_\parallel$, it successfully lines up with the magnetosheath spectrum.  Here only the data that satisfied the selection criteria were retained in the plot.}
%  Note that the interpolation and fitting processes are not directly shown in the figure. }
\end{figure}

To estimate the energization of the sheath electrons, the electron distributions in the sheath are compared to distributions in the solar wind via Liouville mapping. The simplest approach is to only use the field-aligned (or anti-aligned) portion of the particle distributions to conduct the mapping, yielding the quantities $\Phi_\parallel$ and $\Phi_\downarrow$ (which appear, e.g., in Eq.~\ref{delta_phi_eq}). 
%This mapping is appropriate since heating and the magnetic field gradient will only play a minor role in shaping these distributions.
 Therefore, for this part of the analysis only distributions where observations fell within $30^\circ$ of the field are used to represent field-aligned (or anti-aligned) electrons. At low energies photoelectrons can dominate the spectrum, motivating the application of an energy cutoff at 30 eV. The maximum acceleration is expected to be $<$500 eV, which is selected to be the upper energy range for the comparison. Above that energy range, count rates of the instrument are often too low for our purposes and errors in measured strahl can result in an incorrect determination of the energization. % and the solar strahl can result in incorrect weighting.
 So, only energy bins in the 30-500 eV energy range and only energy bins which register $\geq$5 counts are included in the comparison. Examples of two electrons distributions, one from the sheath and a reference spectrum from the solar wind, are presented in Figure~\ref{fit_example_fig}(a). These distributions are from September 22, 2015---the sheath distribution is measured at the nominal time 12h16m55s and the solar wind reference distribution is averaged over a 10-minute period centered on 11h27m44s. 

The solar wind reference distribution, an example of which is presented in Figure~\ref{fit_example_fig}(a), is derived in the same way for each orbit as follows. A time period where the MAVEN spacecraft is located in the solar wind is first identified based on the empirical bow shock position \cite{vignes00}. Then a 10-minute interval is selected, which is centered on the time in the orbit where the satellite is radially farthest from the empirical bow shock position.  The SWEA energy spectra are then averaged over this 10-minute interval to derive one solar wind reference spectrum. For each orbit, \red{all other electron distributions will be compared to this reference spectrum.} 

Note that the electrons measured in the sheath by MAVEN will not generally be found on a flux tube that connects to the position where the solar wind reference spectrum is sampled. It is here assumed that the reference spectrum approximates the source distribution of electrons entering the sheath during a given orbit. 
%This is a reasonable assumption for the parallel and anti-parallel propagating electrons considered here, as observations at Earth have shown that the perpendicular part of the distribution that is more sensitive to the field line connectivity \cite{feldman83}. 
%Also note that because the orientation of MAVEN's orbit with respect to the nominal bow shock varies with the season, MAVEN may never enter the solar wind during certain orbits; such orbits are not included in this study. % not exactly for each 'orbit' note that .

To evaluate whether the field-parallel sheath electrons can be viewed as the solar wind population accelerated by a parallel electric field, a Liouville mapping is performed. To this end the reference spectrum is \red{shifted by a constant energy to best match} each electron distribution of the orbit, yielding the energization ($\Phi_\parallel$ or $\Phi_\downarrow$). For the mapping to be valid, only distributions moving in the same direction with the respect to the magnetic field are compared (i.e. with the same orientation in the solar wind and sheath). However, since the orbital period is long the solar wind magnetic field orientation might change between the times of the sheath and solar wind measurements; say, if the planet encounters a new flux tube in the interim. This fact is accounted for by identifying the orientation of the electron strahl population (if it is significant enough to be identified), which is either aligned or anti-aligned with the magnetic field. The strahl component is required to maintain the same orientation with respect to the field across both the sheath and solar wind in order for the Liouville mapping to be performed---this accounts for some of the natural variability of the interplanetary conditions.
% This way only the direction of the distribution with/without is compared to the same side of the distribution in the solar wind with the observation in the sheath. 
%To illustrate the variation of the electron energy near the bow shock under typical conditions, a study of a single orbit is now presented. Orbit 1908 on September 22, 2015 was selected because the orbital geometry and IMF B angle conditions are such that maximum $\Delta \Phi$ may be expected to be large (i.e. on the order of the expected $\sim$60eV mentioned above). During the interval, the solar wind magnetic field had a cone angle $\theta_c$$\approx$52$^\circ$, which is fairly close to the typical Parker spiral value.

The Liouville mapping is implemented as a least-squares fit, that calculates the energy the solar wind spectrum would need to be shifted by in order to match the sheath spectrum. This energy is denoted as either $\Phi_\parallel$ or $\Phi_\downarrow$, respectively, dependent on whether the fit is conducted between two field-parallel or anti-parallel energy spectra. The results of such a fit for a field-parallel energy spectrum can be seen in Figure \ref{fit_example_fig}(b).
%The Liouville fitting is done in phase space density and the result can be seen in Figure \ref{delta_phi_4plots_fig}2(b) where the sheath distribution has been shifted by $\Phi$ to what it would have been in the solar wind. 
In the example, the derived $\Phi_\parallel$ is 54$\pm$4 eV. The Liouville mapping process involves comparing the energies of two spectra at common values of the phase space density; this comparison is made possible by linearly interpolating the discretely sampled data between the solar wind and sheath spectra.
%Then the energy shift $\Phi_\parallel$ (and its uncertainty) required to match the two spectra are calculated by the least-squared fit.
The fitting is done by weighting each energy bin appropriately by the count rate. The solar wind distribution effectively maps to the sheath distribution at energies $\gtrsim$$\Phi_\parallel$, suggesting the sheath electrons originated from the solar wind. The details of the fitting and the calculation of uncertainties are provided in the supplementary material. %REFERENCE HERE? 

\begin{figure}
\includegraphics[width=1\linewidth]{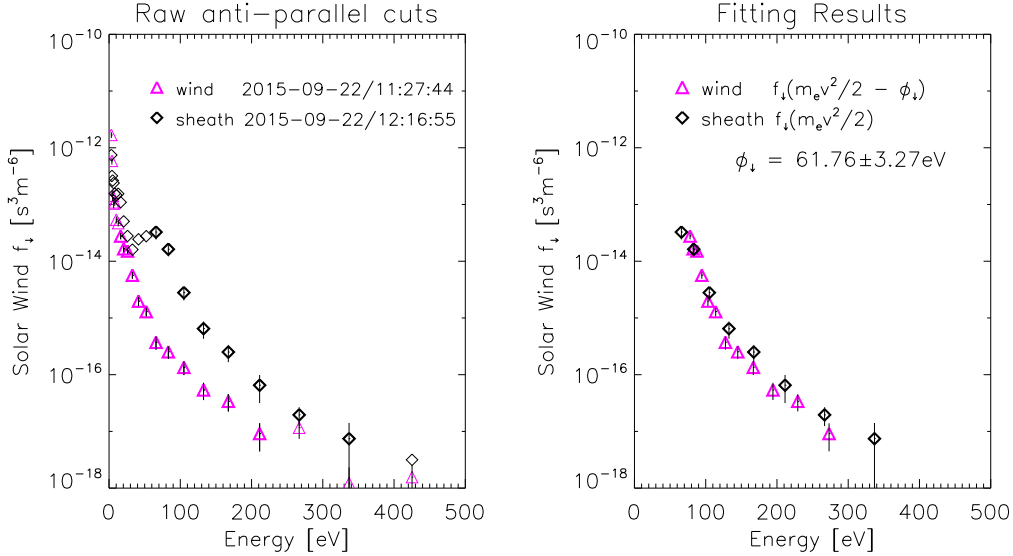}
\caption{\label{fit_example_anti_fig} An example of the antiparallel cuts in the solar wind and sheath, analogous to Figure~\ref{fit_example_fig}. An energization $\Phi_\downarrow$=62$\pm$3 eV is calculated by the fitting procedure. In the right panel, we observe that when the solar wind spectrum is shifted by this energy $\Phi_\downarrow$, it successfully lines up with the magnetosheath spectrum.}
%  Note that the interpolation and fitting processes are not directly shown in the figure. }
\end{figure}

%FIGURE 2
\red{
Figure~\ref{fit_example_anti_fig} shows the same fitting procedure applied to the anti-parallel electrons. These have entered the sheath through the shock at the opposite end of the field line, where the shock potential may be different. The fitting procedure yields $\Phi_\downarrow$=62$\pm$3 eV, similar to the measurement of $\Phi_\parallel$ at the same time.
}

The energy spectra are not corrected for the spacecraft potential $\phi_{sc}$ that arises from spacecraft charging, and this omission introduces systematic error in the estimates of $\Phi_\parallel$ and $\Phi_\downarrow$ of the order $\lesssim$10 eV. In order to more accurately estimate the quantities $\Phi_{\parallel}$, we may want to correct for the spacecraft potential difference, $\Delta \phi_{sc}$, as measured between the locations of the two spectra: for example ${\Phi_\parallel \rightarrow (\Phi_\parallel + \Delta \phi_{sc})}$ and  ${\Phi_\downarrow \rightarrow (\Phi_\downarrow + \Delta \phi_{sc})}$. However, this correction is not of great importance for the present work, as we are interested primarily in measuring the difference $\Delta \Phi$=${(\Phi_\parallel-\Phi_\downarrow)}$, i.e. the spacecraft potential correction cancels out with the subtraction.

%[I would add one of your statistical plot here where all phi is presented as function of Mars Solar Electric (MSE) Ð or choose the layout that is best for your study - and call it Figure 20 in the text below, consider if all IMF B angles should be included or not.]

\begin{figure}
\includegraphics[width=1\linewidth]{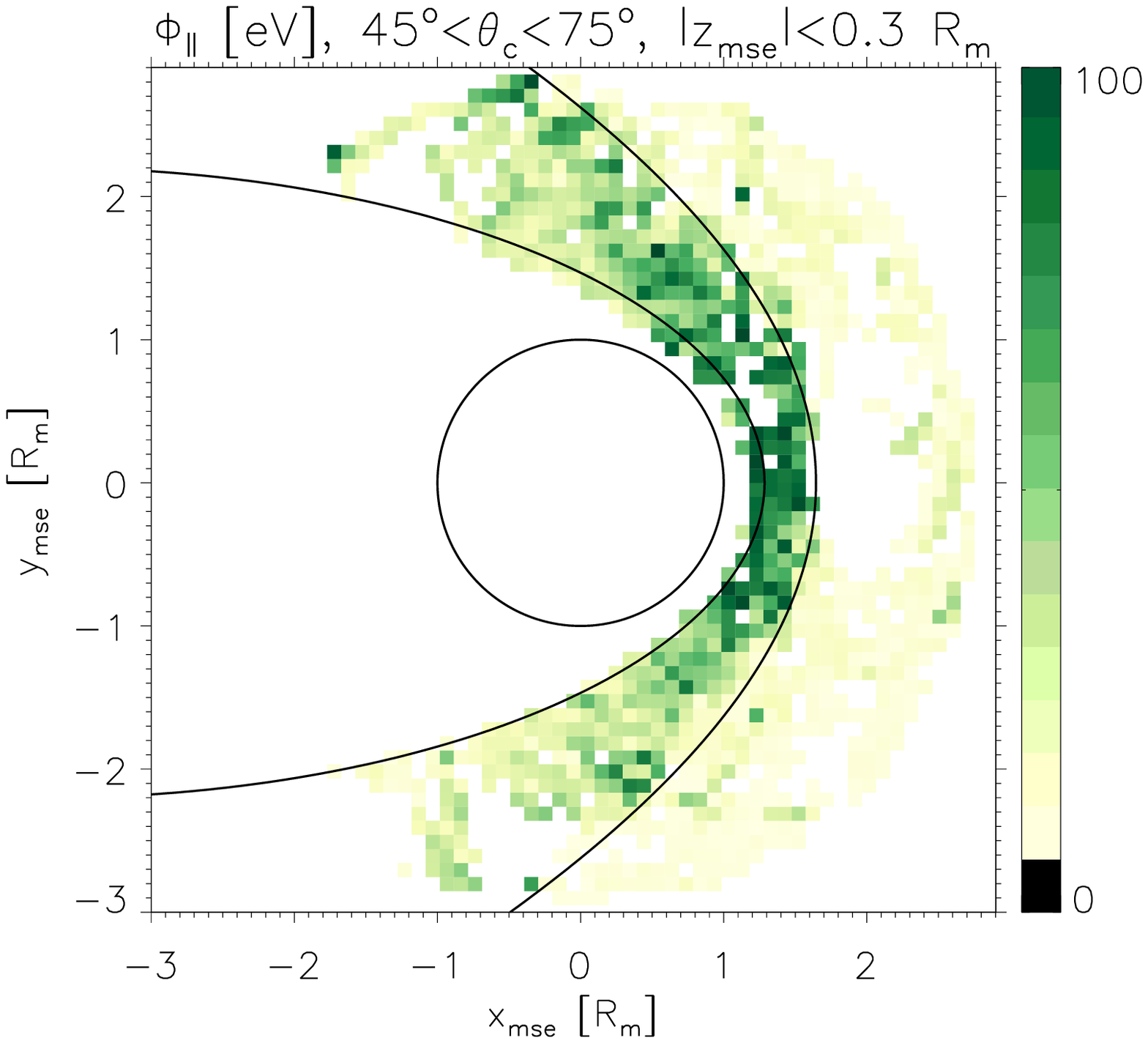}
\caption{\label{phi_par_ave_mse_fig}  Statistical map of the parallel energization $\Phi_\parallel$ averaged across the $>$4-year data set. The data are shown in the Mars Solar Electric (MSE) frame, averaged over times when the spacecraft position satisfied $|z_{mse}|$$<$0.3$R_m$, where $R_m$ is the Martian radius. This plot only includes times when the solar wind magnetic field was directed antisunward, with a cone angle 50$^\circ$$<$$\theta_c$$<$70$^\circ$, as evaluated at the nominal time associated with the solar wind reference spectrum. Overlaid on the plot is the empirical location of the bow shock and MPB \cite{vignes00}.  }
\end{figure}

The energization is calculated via Liouville mapping in this manner for every electron distribution in the $>$4-year data set, enabling the statistical study of $\Phi_\parallel$ and $\Phi_\downarrow$. 
The statistical average of the observed parallel energization $\Phi_\parallel$ is shown in Figure \ref{phi_par_ave_mse_fig}, revealing the spatial structure of the electron energization. 
As expected $\Phi_\parallel$ is nearly zero in the solar wind and increases dramatically near the bow shock location. Near the subsolar point at the bow shock the average $\Phi_\parallel$ is $\sim$100 eV.

%The statistical result of the identified average $\Phi$ as function of Mars Solar Electric (MSE) coordinate is presented in Figure 20 where the color indicates the strength of $\Phi$. Overlaid on the plot is the empirical location of the bow shock [REF]. Only the fits that has an uncertainty of less than 20\% and $\Phi$ is within 30-500 V is included [add all other limitationsÉ.]. Figure 20 show that in the solar wind the natural variability of electron spectrum is within XX V of the reference spectrum. This provide an indication of how well the energy of the sheath electrons can be determined. At the location where the empirical bow shock location is located the value of $\phi jumps to a high (up to ~100 V or less value and inside this location, in the sheath the $\phi is fairly constant). This demonstrate that the main acceleration of the sheath electrons occurs at the bow shock which is expected [REF]. So the dominate parallel acceleration occurs at the bow shock, i.e. a localized acceleration. Secondly, the main acceleration is close to the subsolar point as seen in Figure 20. The magnitude of the acceleration depends on the IMF clock angle and solar wind conditions as have been studies at Earth [REF]. The potential fall is often well described with a cos^2%\theta fall of.

A focus of this study is the physical configuration where a magnetic flux tube is connected at two ends to the bow shock and these two bow shock locations possess different cross-shock potentials. In such a situation the observed electron distributions in the sheath should be asymmetric due to cross-talk. The energies of the two field-aligned electron distributions should show different amounts of acceleration, i.e. one expects $\Phi_\parallel \neq \Phi_\downarrow$. Assuming that the cross-shock potential is roughly cylindrically symmetric about the $x_{mse}$ axis, one expects that the difference between $\Phi_\parallel$ and $\Phi_\downarrow$ would be most suppressed at cone angles $\theta^\circ \approx 90^\circ$. Likewise the effect would be most stark at the smallest angles. From the Parker spiral model one may quickly estimate a typical cone angle of $\theta_c\approx 60^\circ$ at Mars. Under these conditions one expects a maximum difference in energization $\Delta \Phi \approx 60$eV at a location behind the bow shock just offset from the subsolar point (see supplementary material for details).

%To illustrate the variation of the electron energy near the bow shock under typical conditions, a study of a single orbit is now presented. Orbit 1908 on September 22, 2015 was selected because the orbital geometry and IMF B angle conditions are such that maximum $\Delta \Phi$ may be expected to be large (i.e. on the order of the expected $\sim$60eV mentioned above). During the interval, the solar wind magnetic field had a cone angle $\theta_c$$\approx$52$^\circ$, which is fairly close to the typical Parker spiral value.
To illustrate the variation of the electron energy near the bow shock under typical conditions, a study of a single orbit is now presented. Orbit 1907 on September 22, 2015 was selected because the orbital geometry and IMF B angle conditions are such that maximum $\Delta \Phi$ may be expected to be large (i.e. on the order of the expected $\sim$60eV mentioned above). During the interval, the (10-minute avg.) solar wind magnetic field had the value $\bvec{B}$=(-0.93, 2.06, 0.34)nT in MSO cartesian coordinates. This corresponds with a cone angle $\theta_c$$\approx$66$^\circ$, which is \red{within 10\%} to the typical Parker spiral value. This magnetic field is used to calculate the spacecraft vector position in the MSE frame ($\bvec{x_{mse}}$).  The spacecraft's traversal of the sheath takes place over a range of positions satisfying $|z_{mse}|$$\lesssim$0.5$R_m$ ($R_m$ denotes the Martian radius), appropriate for this study since the simple bow shock model assumes ${z_{mse}\approx 0}$ (see supplementary material). Likewise, the spacecraft crosses the shock the near the subsolar point, which is of interest because this is where the strongest signal $\Delta \Phi$ is expected according to the model. \red{Although the exact location of the maximum $\Delta \Phi$ depends on the solar wind cone angle, this expectation may be roughly explained by the fact that in our model the cross-shock potential (which sets $\Phi_\parallel$) peaks at the sub-solar point. This is where the incident ram energy along the vector normal to the bow shock is maximal.}
%That the orbit sample different ages of the flux tubes is of interest since the solar wind flux tubes crossing the bow shock initially is expected to be associated with the largest acceleration and as they are converted in the sheath the potentials decreases and the potential difference between the two bow shock locations also decreases.

\begin{figure}
\includegraphics[width=1\linewidth]{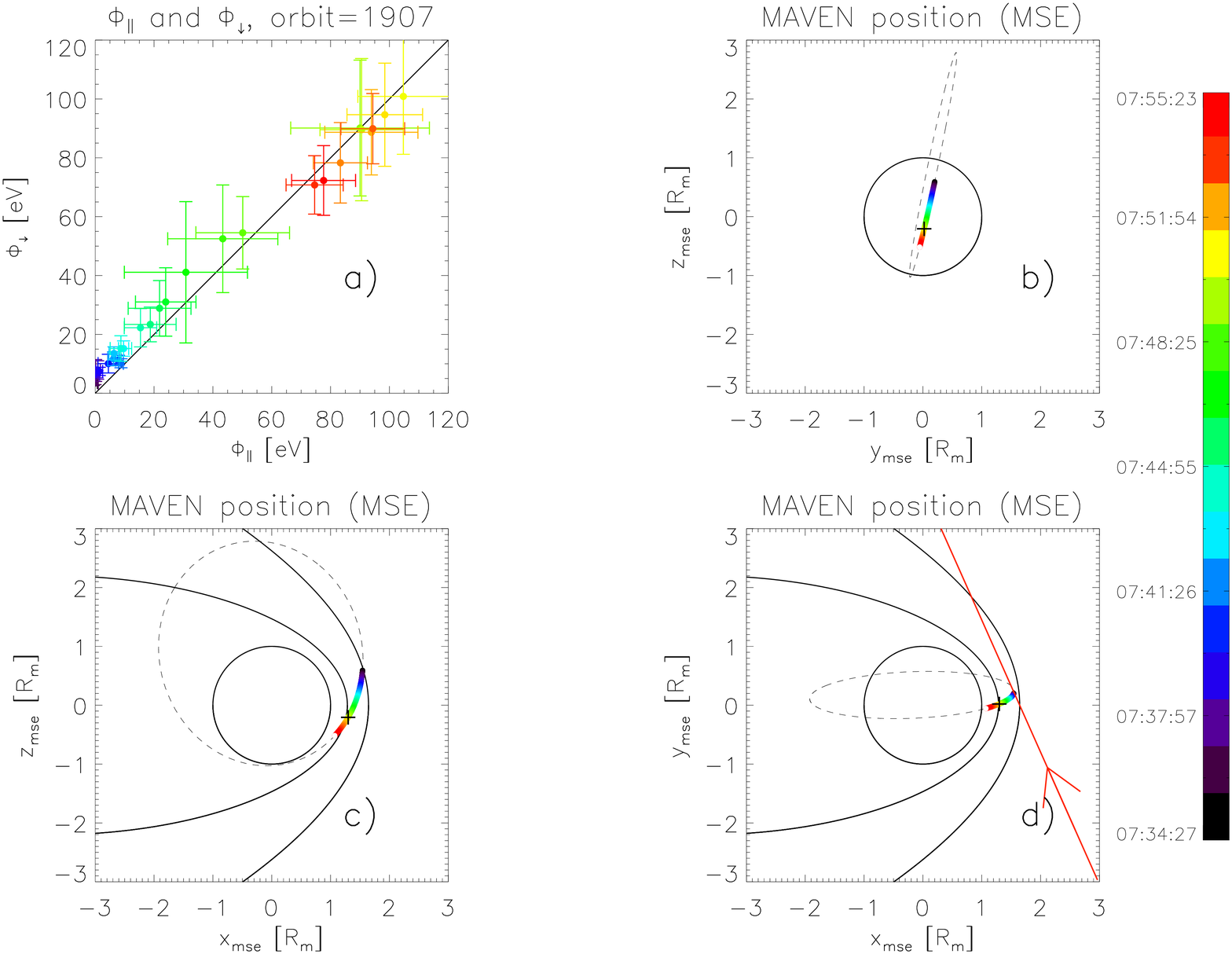}
\caption{\label{phi_anisotropy_1orb_fig}
% Left: A plot of $\Phi_\parallel$ vs. $\Phi_\downarrow$ in the nominal sheath region. Data are from September 22, 2015, during MAVEN's 1907$^{th}$ orbit; in the interval the spacecraft crossed into the sheath near the subsolar point, at $z_{mse}$$\sim$0. Generally it is observed $\Phi_\parallel$$\approx$$\Phi_\downarrow$. Although some small bias $\Delta \Phi$$\lesssim$10eV may be observed at the shock crossing, this signal is much less than that predicted by the model (see text for details). Right: Spacecraft position during the interval, in the $x_{mso}$$-$$z_{mso}$ plane ($|y_{mso}|$$\lesssim$0.1$R_m$). In both plots, the time of day is shown by the color.}
{\bf a.}~A plot of $\Phi_\parallel$ vs. $\Phi_\downarrow$ in the nominal sheath region. Data are from September 22, 2015, during MAVEN's 1907$^{th}$ orbit; in the interval the spacecraft crossed into the sheath near the subsolar point, at $z_{mse}$$\sim$0. Generally it is observed $\Phi_\parallel$$\approx$$\Phi_\downarrow$. Although some small bias $\Delta \Phi$$\lesssim$10eV may be observed at the shock crossing, this signal is much less than that predicted by the model (see text for details).  {\bf b.}~Spacecraft position in the $y_{mse}$$-$$z_{mse}$ plane.  {\bf c.}~Spacecraft position in the $x_{mse}$$-$$z_{mse}$ plane. {\bf d.}~Spacecraft position in the $x_{mse}$$-$$y_{mse}$ plane. As the magnetic field in MSE coordinates falls exactly in this plane, a line showing the solar wind magnetic field orientation during the interval ($\theta_c$=66$^\circ$) is shown for reference. In all plots (a)-(d), the time of day is shown by the color. The Martian surface, MPB, and bow shock boundary are shown as solid lines where applicable in plots (b)-(d), and MAVEN's orbital trajectory is shown as a dashed line.}
\end{figure}

The results of the fits from MAVEN's 1907$^{th}$ orbit on September 22, 2015 are presented in Figure~\ref{phi_anisotropy_1orb_fig}(a). The $\sim$30-minute time interval during which the spacecraft crossed into the sheath is divided into 30 subintervals of $\sim$1-minute duration, and averages of $\Phi_\parallel$ and $\Phi_\downarrow$ within those subintervals are plotted. The central time within each subinterval is displayed by the color. The standard deviation of the $\Phi_\parallel$ and $\Phi_\downarrow$ data within each subinterval (the scatter) is displayed as error bars.
%For each time measured by SWEA, the fitted energies $\Phi_\parallel$ and $\Phi_\downarrow$ are marked by a dot which is color-coded by the time within the orbit. 
Recall the quantities  $\Phi_\parallel$ and $\Phi_\downarrow$ represent the derived net energization that the two electron populations (moving along and against the field) have each experienced.
The spacecraft location for the selected orbit is presented in Figures~\ref{phi_anisotropy_1orb_fig}(b)-(d) with the same color coding. 
%This orbit is close to Zmso /approc 0 and the solar wind magnetic field had an cone angle $\theta_c$$\approx$52$^\circ$. % [note typo in the plot either everything is mso or everything is mse].  

As discussed above, for the conditions of the selected orbit one may expect $\Delta \Phi$$ \sim $60~eV just downstream of the bow shock. But, $\Phi_\parallel$ and $\Phi_\downarrow$ for this orbit lie along the unity line (solid line) in Fig.~\ref{phi_anisotropy_1orb_fig}(a), suggesting that the magnitudes of the energizations are actually quite similar. The time where the spacecraft crosses the bow shock (07h49m20s) has been identified by manually looking at the data and is marked by the ``+'' sign in Figure \ref{phi_anisotropy_1orb_fig}. Note that this shock crossing is detected at a lower altitude than that of the average bow shock \cite{vignes00}, indicating that the shock happened to be relatively compressed during this orbit. The largest $\Delta \Phi$ should be observed just behind the bow shock, near the subsolar point. But no such systematic difference is observed, as the points in Fig.~\ref{phi_anisotropy_1orb_fig}(a) adhere to the unity line throughout the time interval.
% The scatter in the data is slightly larger at the bow shock (where $\Phi_\parallel$, $\Phi_\downarrow$$\sim$100 eV), but a clear trend suggesting either $\Phi_\parallel$ or $\Phi_\downarrow$ is larger cannot be seen. 
The scatter in the data is slightly larger at the bow shock (where $\Phi_\parallel$, $\Phi_\downarrow$$\sim$100 eV). Although there seems to be a slight bias $\Phi_\parallel$$>$$\Phi_\downarrow$ at this time, the magnitude of $\Delta \Phi$ is only $\sim$10 eV (i.e. $\ll$60 eV).
This figure therefore suggests that electrons in the sheath cannot have been accelerated at the bow shock alone.

%`As the magnetic fields tubes comes closer to the planet the magnitude of \phi will decrease which Figure 3. Flux tubes closer to the planet is associated with lower \phiÕs which is suggested based on Figure 20, since these flux tubes maps to higher clock angles. The scattering observed at these lower \phiÕs is slightly less than the younger flux tubes with higher \phiÕs. 

\begin{figure}
\hspace{-4cm}\includegraphics[width=1.5\linewidth]{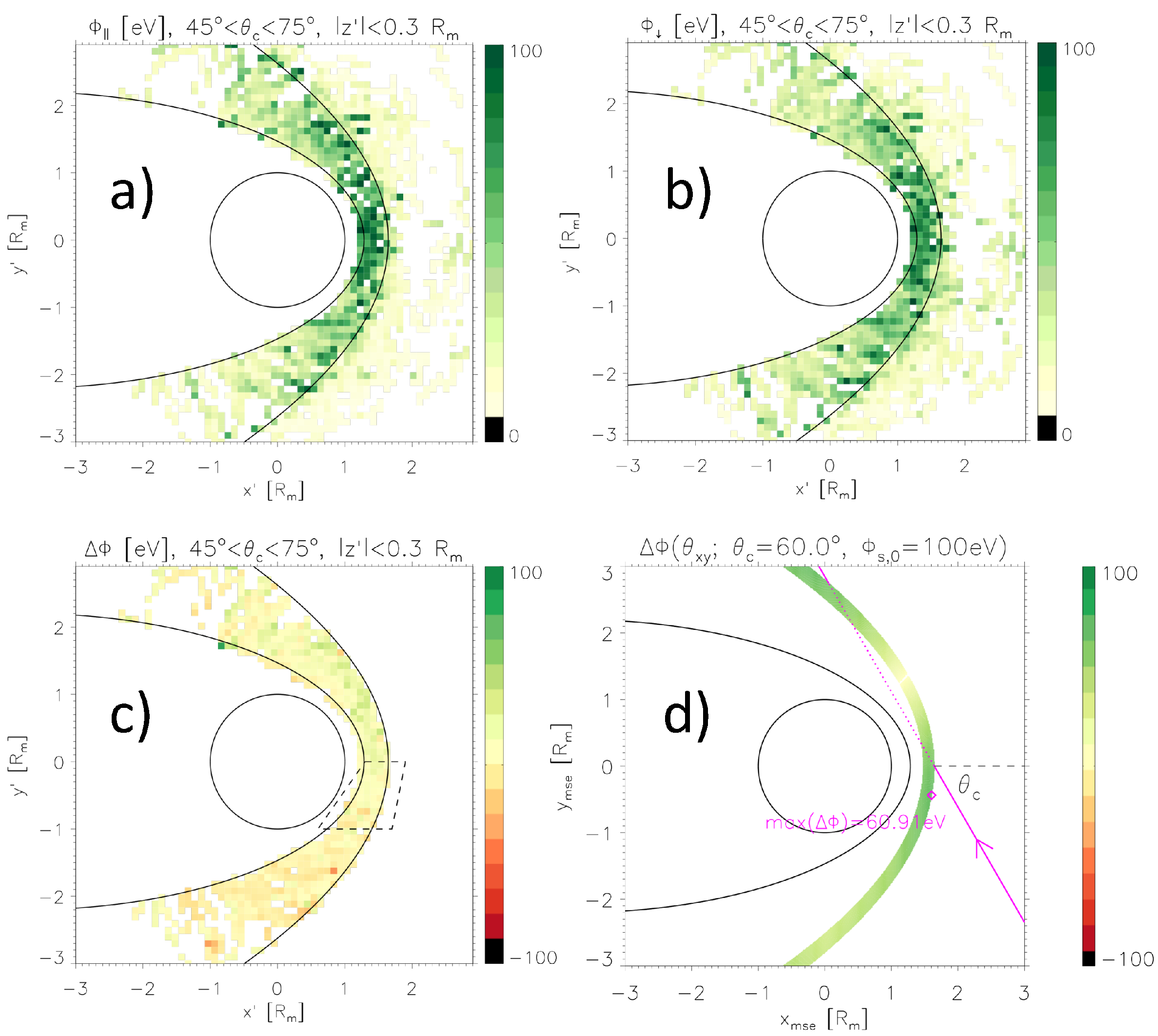}
\caption{\label{delta_phi_4plots_fig} \red{The electron energization in the Martian magnetosheath, in the $z^\prime$=$z_{mse}$=0 plane under typical conditions ($\theta_c$=60$^\circ$).} {\bf a.}~Average parallel energization, $\Phi_\parallel$, plotted in the shock-normalized MSE frame. Before averaging the data, we normalize distances to the inferred scale of the shock (see text). We further restrict ourselves to data satisfying $|z^\prime|$$<$0.3$R_m$, and only consider data when the solar wind cone angle fell within the range 50$^\circ$$<$$\theta_c$$<$70$^\circ$. {\bf b.}~Same as (a), but here we plot the average anti-parallel energization $\Phi_\downarrow$. {\bf c.}~Same as (a), but here we plot the energization difference ${\Delta \Phi=\Phi_\parallel-\Phi_\downarrow}$. \red{Only data measured within the magnetosheath are included, to highlight the variation in this region.} {\bf d.}~The predicted $\Delta \Phi$, that arises from a model in which the electron energization occurs solely at the bow shock. The model assumes $\theta_c$=60$^\circ$ (the typical cone angle observed at Mars) and $z^\prime$=0, so it may be compared with (c). \red{A magnetic field line with cone angle $\theta_c$=60$^\circ$ is shown for reference, but note that in the sheath region (dashed lines) the field is expected to be curved, not straight.} Note the dissimilarity between the $\Delta \Phi$ signature in (c) and (d)---see text.}
\end{figure}

To see if the observed trend $\Phi_\parallel\approx \Phi_\downarrow$ holds generally for other orbits, a statistical evaluation of the two energies $\Phi_\parallel$, $\Phi_\downarrow$ and their difference $\Delta \Phi$ is presented in Figures~\ref{delta_phi_4plots_fig}(a)-(c). Again only times for which the cone angle satisfied 50$^\circ$$<$$\theta_c$$<$70$^\circ$ are considered in the averages.
 The individual $\Phi_\parallel$ and $\Phi_\downarrow$ values that go into the averages are calculated as already described in this section (following the same selection criteria). 
The $>$4 years of data are aggregated by calculating the spatial averages of these quantities in a dynamic coordinate system. \comment{reference?} % (similar to what has been done for the Earth's auroral oval, W. K. Peterson et al). \comment{reference?}

%Only electrons within 30$^\circ$ of the magnetic field are used to represent the field aligned electrons spectrum. Times when the magnetic field is not within SWEA's field of view are omitted from the statistics. 

The dynamical coordinate system is developed as follows. For each orbit, the location of the bow shock is identified from among times where the derived quantity $\overline \Phi \equiv (\Phi_\parallel + \Phi_\downarrow)/2$ is in the 99$^{th}$ percentile for that orbit. This simple criterion is used because the electrons are known to be strongly energized at the shock. From among these times, the shock crossing is designated as the location where the spacecraft altitude is maximal. Once the location of the bow shock has been specified, this information is used to estimate the local scale of the bow shock relative to the nominal shock size \cite{vignes00}. The spacecraft's position in MSE coordinates is normalized by the local shock size (evaluated each orbit) before conducting the spatial averages presented in Figures~\ref{delta_phi_4plots_fig}(a)-(c). These normalized MSE positions are denoted by the vector components $x^\prime$, $y^\prime$, $z^\prime$. Normalizing in this way minimizes the effects of natural variance of the system. For instance, the signature of the electron energization near the shock is less blurred out by the time-varying size of the shock, and the sheath and solar wind populations are well-separated before averaging.

The presented statistical maps should only include data with similar cone angle $\theta_c$, so as not to not mix electrons originating from different locations along the bow shock. The typical $\theta_c$ at Mars is about 60$^\circ$, as predicted by the Parker spiral model. For Figure~\ref{delta_phi_4plots_fig} we therefore only include data where the cone angle at the nominal time of the solar wind reference spectrum's measurement satisfied 50$^\circ$$<$$\theta_c$$<$70$^\circ$. Analogous, nearly identical plots  (not shown) may be produced for the cases when the magnetic field had the same orientation with opposite polarity ($\theta_c$=120$^\circ$).

In the averages presented in Figure~\ref{delta_phi_4plots_fig}, only data from the regions nominally occupied by the solar wind and sheath are included \cite{vignes00}. Also, only data where the spacecraft position satisfied $|z_{mse}|$$<$0.3$R_m$ are included. This reflects the fact that the parallel asymmetry, if it exists, should be most stark in the plane $z_{mse}$=0. The (unscaled) surface of Mars and the empirical boundary locations of the bow shock and the transition region where the MPB is located are presented by the black lines in the figure. Also, a drawing of the magnetic field with  $\theta_c$=60$^\circ$ illustrates how the magnetic field encounters the system; this is shown as a dashed line in the sheath region because the draped field is actually curved there. It is quite obvious that the statistical result of Figures~\ref{delta_phi_4plots_fig}(a)-(b) is similar to Figure~\ref{phi_anisotropy_1orb_fig}---that is, $\Phi_\parallel$ and $\Phi_\downarrow$ are similar in magnitude and the largest $\Phi$ are seen close to the subsolar point.

Because of the selected range of cone angles, $50^\circ$$ <$$ \theta_c$$<$$70^\circ$, the energization of the electron distributions may be expected to be asymmetric. That is, we may expect ${\Phi_\parallel \neq \Phi_\downarrow}$, and some difference between Figure~\ref{delta_phi_4plots_fig}(a) and~\ref{delta_phi_4plots_fig}(b) should be observed. 
%However it is not as large as expected but could be due to averaging a data set that has large variability in it. 
Therefore $\Delta \Phi$ is first calculated for each individual point before deriving the average, which is presented in Figure~\ref{delta_phi_4plots_fig}(c).
 The difference is close to zero throughout the sheath, which is an unexpected result in this study. Near the bow shock there is a slight trend in the $\Delta \Phi$ data, with $\Delta \Phi \gtrsim 0$ in the region $y_{mse}>0$ and  $\Delta \Phi \lesssim 0$ in the region $y_{mse}<0$. This systematic signal is strongest at the flanks, with a maximum strength of 10-20 eV. The region where the maximum $\Delta \Phi$ may be expected is outlined by a dashed trapezoid---although the model predicts $\Delta \Phi$$\sim$60eV in this region (see next paragraph), the actual signal varies within the range -21eV$<$$\Delta \Phi$$<$16eV. In other words, the observed $\Delta \Phi$ is only about $\lesssim$25\% of the predicted value in the region where the signal is expected to be strongest.  %That is, we generally see $\Delta \Phi >0$ at positions $y^\prime>0$ and $\Delta \Phi <0$ at positions  $y^\prime<0$]. This asymmetric trend is expected but were expected to be larger. 

For comparison, Figure~\ref{delta_phi_4plots_fig}(d) shows a model prediction of $\Delta \Phi$ just downstream of the bow shock in the region $z_{mse}$=0, assuming a cone angle of 60$^\circ$ and a peak cross-shock potential of 100 eV at the subsolar point (see supplementary material for details). For such conditions the expected $\Delta \Phi$ is estimated to be 61 eV, as calculated from the difference in cross-shock potentials at two ends of a flux tube.  In the model, ${\Delta \Phi=0}$ near where the downstream magnetic field is tangent to the shock surface ${\bvec{B} \cdot \hat \bvec{n} }$=0.

% The potential difference presented in Figure~\ref{delta_phi_4plots_fig} is what was expected and this expected value is much larger than what the data in panel c shows. 

The predicted signal  Figure~\ref{delta_phi_4plots_fig}(d) differs from the observed signal Figure~\ref{delta_phi_4plots_fig}(c) in a number of important respects. Note that the model predicts $\Delta \Phi$$\geq$0 throughout the sheath, whereas the actual signal skews negative in the region $y_{mse}<0$, as mentioned above. Also the observed $\Delta \Phi$ is very close to zero near the subsolar point, whereas this is where the model predicts $\Delta \Phi \approx $60 eV. In fact, the observations show no systematic signal of strength $\sim$60 eV anywhere in the sheath, and in the statistical map the \red{maximum and average values of $|\Delta \Phi|$ observed  are about 20~eV and 8~eV respectively}.

%This estimation is accomplished by considering the average energization $\overline\Phi \equiv (\Phi_\parallel +\Phi_\downarrow )/2$ at every time this quantity was measured during one orbit. [I do not understand thisÉ.. why 2??  Is this the correct way to identify the boundary??]

We have assumed so far that the electrons in the sheath have evolved collisionlessly, and moreover that the magnetic moment is a conserved quantity. If other processes such as collisions and nonlinear wave acceleration are present in the flux tube, the modification of the distributions will be clearly observable in the particles with large pitch angles. Therefore an investigation is made to see if the angular distributions of sheath electrons can be described as solar wind electron distributions that have been exposed to a quasi-static electric field (ignoring nonlinear effects).  In investigating electrons with large pitch angles, magnetic field gradients must also be considered. This investigation is based on Liouville mapping; first the field-aligned distribution is used to identify the energization (parametrized by $\Phi_\parallel$, $\Phi_\downarrow$) due to the electric field and then conservation of magnetic moment is applied to calculate the mapped angular distribution. Note that in the presence of a spatially-varying magnetic field, the pitch angles of particles will change in order to conserve the magnetic moment. %%[write the simple equation B_1/B_2/=PA_1/PA_2]. 

%If other processes such as collision or nonlinear wave acceleration is present on the flux tube the modification of the distributions would be more observable on the particles with small parallel speed to the magnetic field, i.e. large pitch angles. In investigate large pitch angle electrons magnetic field gradients has to be also taken into consideration. Therefore an investigation will be made to see if the sheath electrons can be describes as a solar wind electron distribution has been exposed to a localized quasi-static potential, i.e the bow shock, followed with a magnetic field gradient. Again this will be done based on Liouville mapping by fist using the field aligned distribution to identify the potential acceleration and apply the changes based on the magnetic field. The magnetic field gradient will result in an mirror force, i,e, to conserve the magnetic moment the pitch angle will change based on the gradient without losing any kinetic energy %%[write the simple equation B_1/B_2/=PA_1/PA_2]. 

\begin{figure}
\includegraphics[width=1\linewidth]{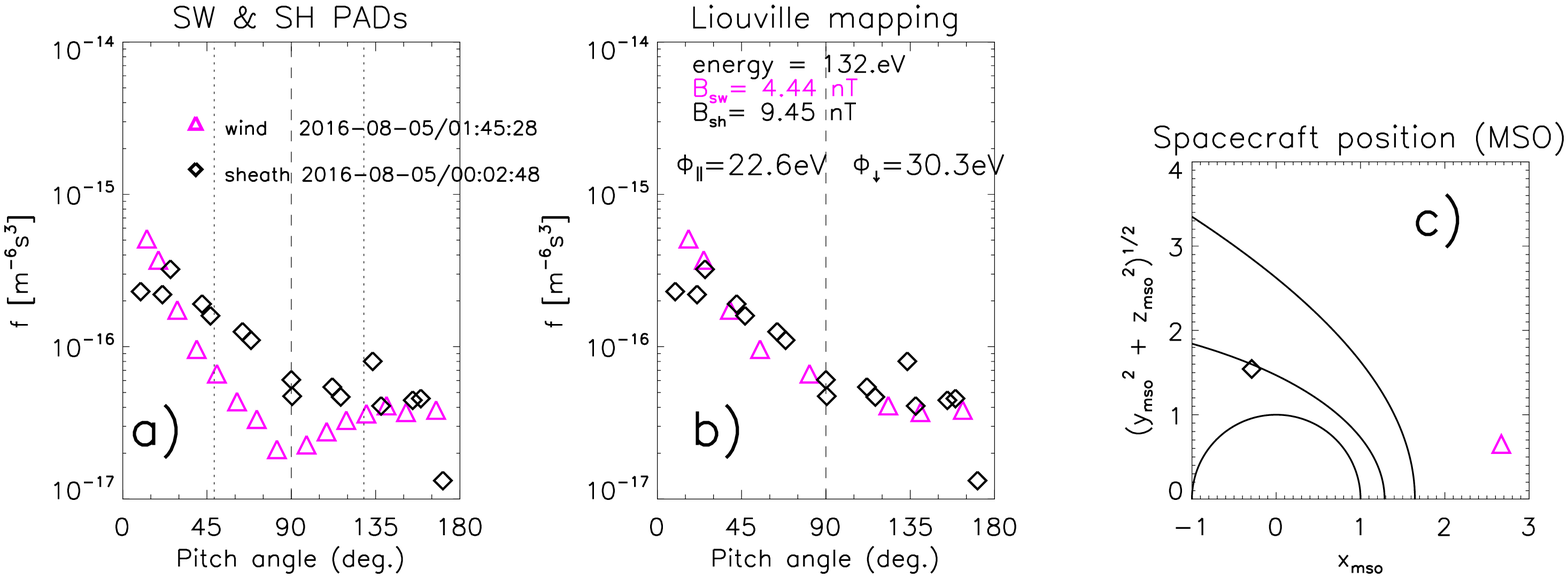}
\caption{\label{pad_mapping_fig} Liouville mapping of pitch angle distributions for a particular case study, \red{chosen because the solar wind distribution exhibited a prominent strahl population during the orbit.} {\bf a.}~Sheath pitch angle distribution measured at 132eV, and a solar wind pitch angle distribution interpolated at energies $\sim$102-110eV (appropriate for Liouville mapping in pitch angle, see text). {\bf b.}~Results of the Liouville mapping. The sheath distribution is shown again as in panel a), to be compared with the Liouville mapped solar wind distribution. The curves line up well, indicating a successful mapping. {\bf c.}~Position of the spacecraft at the time of measurement, respectively for the solar wind (triangle) and sheath (diamond) distributions. \red{The average position of the bow shock and MPB \citeA{vignes00} are shown for reference.}}
%[Is the units provided in the figure correct?]
\end{figure}

Two pitch angle distributions, one from the solar wind and another from the sheath, are presented in Figure~\ref{pad_mapping_fig}(a). The data are from 5 August, 2016 where the sheath was measured at 00h02m48s and the solar wind reference spectrum is derived from a 10-minute window centered at 01h45m28s.  The sheath distribution was measured by SWEA at energy 132 eV.
%and the solar wind distribution is interpolated to the same energy, as will be described momentarily. 
The solar wind pitch angle distribution displayed in Fig.~\ref{pad_mapping_fig}(a) is taken from the solar wind reference as usual, but interpolated to the appropriate energy. That is, the field-aligned part ($\theta$$<$90$^\circ$) of the solar wind distribution is interpolated to the energy $\mathcal{E}_\parallel$, so that the total energy of the electrons (after they migrate from the solar wind to the sheath location) would match the sheath distribution; i.e., $\mathcal{E}_\parallel+\Phi_\parallel$=132 eV. An analogous interpolation process, using $\Phi_\downarrow$, is used to construct the rest of the distribution (where $\theta$$>$90$^\circ$).  In this way, a solar wind pitch angle distribution is constructed that may be mapped to the sheath distribution. 

%Based on the instantaneous magnetic field strength at the two location (9.5 vs 4.6 nT) for which the magnetic field gradient is assumed to only occur inside the bow shock. The fitting suggest $\Phi$ is 22 V and the solar wind distribution is mapped into the sheath and presented together with the sheath distribution in Figure~\ref{pad_mapping_fig}(b). [I have missed something because you are mapping from two direction so I should have a ($\Phi_\parallel$  and $\Phi_\downarrow$ but I only find $\Phi$]

%

The results of the Liouville mapping are presented in Fig.~\ref{pad_mapping_fig}(b). In this panel the solar wind distribution from panel \ref{pad_mapping_fig}(a) has been mapped to new pitch angles---as the distribution would appear once the electrons had migrated to the selected location in the sheath. The change in pitch angle $\theta$ can be predicted by accounting for the gain in parallel energy ($\Phi_\parallel$ and $\Phi_\downarrow$, derived as shown in Fig.~\ref{fit_example_fig}) and the conservation of magnetic moment. That is, given an electron's initial velocity components $v_{\parallel,1}$, $v_{\perp,1}$, the ratio of the magnetic fields in sheath and solar wind $B_{sw}/B_{sh}$, and the gain in energy $\Phi_\parallel$ (or $\Phi_\downarrow$), the final velocity of a particle can be computed. The relevant equations for computing the final (sheath) velocity components are described in section~1.1 of the supplementary document.

For the purposes of the pitch angle mapping, the magnetic field values in the solar wind and sheath locations were measured as $B_{sw}$=4.6 nT and $B_{sh}$=9.5 nT. The parallel and anti-parallel energization of the solar wind electrons were $\Phi_\parallel$=22.6eV and $\Phi_\downarrow$=30.3eV, respectively. \red{The position of the MAVEN spacecraft at the two locations, in MSO coordinates, is shown in Figure~\ref{pad_mapping_fig}(c).}

%Since the acceleration of the electrons shifts the distribution to a new energy only looking at the same energy in the two distributions in Figure~\ref{pad_mapping_fig} is not sufficient. Therefore the mapping was made with the simple approach first fit the parallel distribution to find the acceleration (as was done in Figure 2). Then applied this acceleration to the solar wind distribution so that the same energy can be compare between the solar wind and the sheath. Thereafter apply the mirror force based on the observed gradient. [Is this what you did or did you fit the full distribution in both B and E at the same time and then selected what to present in Figure 7a?]

\red{For this example,} no mapping is conducted at pitch angles $|\theta$$-$$90^\circ| $$\lesssim $$40^\circ$ because the ratio of magnetic fields dictates that any such particles from the solar wind would have been reflected (the mirror condition) before they reached the location of the sheath observation. The pitch angles corresponding to the mirror condition, for the particular magnetic fields used in the mapping, are shown as vertical lines in Fig.~\ref{pad_mapping_fig}(a). Note that these boundaries are only valid if a strong magnetic field doesn't exist between the sheath and solar wind locations---such a field would change the domains in phase space of the ``passing'' and ``reflected'' populations.

Figure~\ref{pad_mapping_fig}(b) demonstrates that there is good agreement between the solar wind reference distribution and the Liouville-mapped sheath distribution.
%For the sheath distribution portion that maps out to the solar wind the agreement in Figure~\ref{pad_mapping_fig}(b) with the solar wind data is good. 
%The solar wind is not the source to the particles close to 90\o. These particles originates form the sheath electrons by instabilities and waves. Field aligned accelerated distribution is expected to become more isotropic in pitch angle if pitch angle scattering or perpendicular wave heating are present.
If pitch angle scattering or perpendicular wave heating were present in the sheath, the Liouville mapping would fail and the sheath distributions would be expected to be more isotropic.
 The importance of these two processes is small, as can be seen from the significant asymmetry of the distribution, where the phase space density of the field-parallel beam ($\theta$=$0^\circ$) is about an order of magnitude greater than that of the \red{anti-parallel} electrons. 
This inferred lack of pitch-angle scattering provides a key to understanding the similarity between $\Phi_\parallel$ and $\Phi_\downarrow$ as presented in Figure~\ref{delta_phi_4plots_fig}.
%The explanation of Figure~\ref{delta_phi_4plots_fig} cannot for instance be mirroring or pitch angle scattering of particles, that may result in a 180$^\circ$ rotation of the electron velocities. If these two processes were important the fluxes in parallel and perpendicular direction would be similar. % [Can you make a statistical plot or something supporting this statement??? I think such a plot would help you allot with your theory presented in the discussion] 

\section{Discussion}\label{discussion_sec}

In this study we have operated under the assumption that the Martian magnetosheath is a collisionless medium. We have further assumed that 1) the electrons observed in the sheath are sourced from the solar wind and are only energized by a parallel (or anti-parallel) electric field that increases the particle energy by an amount $\Phi_\parallel$ (or $\Phi_\downarrow$), and 2) the solar wind and sheath do not vary dramatically on timescales $\Delta t$$\sim$1 hour, allowing distributions measured at two points in the MAVEN orbit to be compared. This simple framework is sufficient to explain the field-aligned sheath distributions (e.g., Fig.~\ref{fit_example_fig}), and accounting for the magnetic field can also explain the pitch angle distributions (Fig.~\ref{pad_mapping_fig}).
%When taken into account for potential magnetic fields gradient also fluxes at pitch angles can be explained as shown in Figure 7.
Therefore any process of collisions or plasma wave acceleration must have only a minor effect on the sheath electron distributions. 

%The distribution of the field aligned-acceleration, i.e. $\Phi$, is dominated by the acceleration at the bow shock as can be seen in Figure~\ref{phi_anisotropy_1orb_fig}.  

%Above we have provided two scenarios above that may account for the relatively small values of $\Delta \Phi$ presented in section~\ref{obs_map_sec}. 

This study shows that the electron energization is more symmetric than expected from the electron ``cross talk'' picture, as demonstrated by Figures~\ref{phi_anisotropy_1orb_fig} and~\ref{delta_phi_4plots_fig}. This suggests that the model of local shock energization, quantified explicitly in the supplementary document, does not fully describe the electron distributions. Some additional acceleration must be introduced to explain the relative symmetry of the energization experienced by the sheath electrons. This leads us to consider two different explanations for why the observed $\Delta \Phi$ is not more significant, detailed below.
%, or more specifically whether electron acceleration may be occurring within the sheath itself.

\subsection{Scalar potential $\phi(\bvec{x})$}\label{discussion_pot_sec}

One explanation for the observation $\Delta \Phi$$\sim$0 may be that the electrons are not exclusively accelerated at the shock front as we originally assumed.  Rather, the particles may also respond to parallel electric fields as they traverse the magnetosheath.  We stress that rapid energization would still occur as the electrons cross the bow shock, but the electrons would also be gradually accelerated by the presumed electric field in the sheath. 

%In this interpretation, we may express the electron energization $\Phi(\bvec{x})$ by the familiar formula, simply the conservation of (kinetic plus potential) energy:

%\begin{equation}\label{phi_electrostatic_eq}
%\Phi(\bvec{x}) = |q_e| \phi(\bvec{x}).
%\end{equation}

%\noindent Note the absolute value in Eq.~(\ref{phi_electrostatic_eq}) is included to conform with the sign conventions used, e.g., in Eq.~(\ref{v_par_eq}) and in the rest of the paper---empirically, electrons gain kinetic energy when crossing the shock into the sheath, so $\Phi\geq 0$.

%Attributing the electron energization to a spatially varying electrostatic potential (Eq.~\ref{phi_electrostatic_eq}) may neatly resolve the issue of electron anisotropy. In this picture, all electrons arriving at position $\bvec{x}$ in the sheath would have gained the same energy $\Phi(\bvec{x})$---regardless of the location at which they crossed the bow shock into the sheath---so throughout the system we would expect $\Delta \Phi $=0. 

%We stress that rapid energization would still occur as the electrons cross the bow shock, but the electrons would also be gradually accelerated by the electric field as they traversed the sheath.
% The ambipolar electric field that arises from an electron pressure gradient could provide the necessary acceleration. 

In this interpretation, the observation $\Delta \Phi$$\approx$$0$ suggests that the energizing electric field is itself the gradient of some scalar potential $\phi(\bvec{x})$ in the shock and magnetosheath. That is, the energization of an electron depends on its position rather than its trajectory. In this view, we may drop the subscripts $\parallel$, $\downarrow$ and express the energization $\Phi$ as a sole function of the position $\bvec{x}$:

\begin{equation}\label{phi_electrostatic_eq}
\Phi(\bvec{x}) = |q_e| \phi(\bvec{x}).
\end{equation}

% In this interpretation, electrons would  we may express the electron energization $\Phi(\bvec{x})$ by the familiar formula, simply the conservation of (kinetic plus potential) energy:

\noindent Note that $q_e$$<$0 is the electron charge in Eq.~(\ref{phi_electrostatic_eq})---empirically, electrons gain kinetic energy as they cross into the sheath, so $\phi$$\geq$0.
Such a picture may neatly account for the lack of electron anisotropy that would otherwise be expected from cross-talk.

To explain the electrostatic potential we will invoke the presence of an ambipolar electric field $\bvec{E}_A$. Such fields are found in the presence of electron temperature and density gradients, and such gradients may be seen in the quasi-steady sheath. Indeed, the cross-shock energization $\Phi_s$ (see supplementary material) is widely attributed to the ambipolar field established in the Martian bow shock, where these gradients are most pronounced. A weaker ambipolar field in the sheath region could still have a significant effect on the electron energy, as the total distance traveled by an electron through the sheath is much greater than the cross-shock distance. 

Let us investigate the effect of $\bvec{E}_A$ on the electron energization in the Martian frame, by considering the following formula for the electric field, which follows from retaining the leading terms of the steady-state electron momentum equation (ignoring the momentum transport term):

\begin{equation}\label{E_ambi_eq}
%\bvec{E}(\bvec{x}) = \frac{1}{n_e q_e}\nabla P -\bvec{v}_{b}\times\bvec{B},
\bvec{E}(\bvec{x}) = \bvec{E}_C(\bvec{x}) + \bvec{E}_A(\bvec{x}).
\end{equation}

%\noindent where $\bvec{E}(\bvec{x})$ is the static electric field, $n_e$ is the electron density, $P=n_eT_e$ is the electron pressure (assuming isotropic electrons with kinetic temperature $T_e$), and $\bvec{v}_{b}$ is the electron bulk flow velocity. We also introduced the notation $\bvec{E}_a$, which denotes the ``ambipolar'' electric field:
%\noindent where $\bvec{E}(\bvec{x})$ is the static electric field, $\bvec{v}_{b}$ is the electron bulk flow velocity, and $\bvec{E}_a$ denotes the ambipolar field:
\noindent In Eq.~\ref{E_ambi_eq}, the convective ($\bvec{E}_C$) and ambipolar ($\bvec{E}_A$) contributions to the electric field are given by the standard expressions:

\begin{eqnarray}\label{E_conv_def}
\bvec{E}_C \equiv  -\bvec{v}_{b}\times\bvec{B},\\
\label{E_ambi_def}
\bvec{E}_A \equiv \frac{1}{n_e q_e}\nabla P_e.
\end{eqnarray}

\noindent In Eq.~(\ref{E_conv_def}-\ref{E_ambi_def}), $\bvec{v}_b$ is the bulk flow velocity, $n_e$ is the electron density and $P_e=n_eT_e$ is the electron pressure (assuming isotropic electrons with kinetic temperature $T_e$). Assuming $T_e$ varies by about 100eV over the magnetosheath scale $\sim$$10^7$m, we may estimate the typical magnitude of the ambipolar field: $|\bvec{E}_A| \sim 10^{-5}$V/m. Assuming typical values in the sheath $|\bvec{v}_b|$$\sim$$10^5$m/sec and $|\bvec{B}|$$\sim$$10^{-8}$T, we find $|\bvec{E}_C|$$\sim$$10^{-3}$V/m, i.e. the ambipolar field is about 1/100 the typical convective field.
%This is about 1/100 the typical convective electric field in the sheath $|\bvec{E}_C|\sim 10^{-3}$V/m.
 We note however, that this estimated ambipolar field (ignoring the convective field) may alone account for the observed electron energization $\Phi\sim$100eV when integrated across the magnetosheath scale. Indeed, to a first approximation the convective term in Eq.~(\ref{E_ambi_eq}) may be ignored for the purpose of understanding electron energization---as will be shown shortly.

%The pressure term in Eq.~(\ref{E_ambi_eq}) gives rise to a so-called ``ambipolar'' electric field. \comment{cite other related studies of ambipolar field? Or Lathys hybrid simulations?}

Let us consider the trajectory of an electron that moves through the electric field (\ref{E_ambi_eq}) with guiding center velocity $\bvec{v}_{gc}$: % that includes the ${\bvec{E}\times\bvec{B}}$ drift:
\begin{equation}\label{v_gc_eq}
%\bvec{v}_{gc} = v_{gc,\parallel} \hat \bvec{b} + \frac{\bvec{E} \times \bvec{B}}{B^2},
\bvec{v}_{gc} = v_{\parallel} \hat \bvec{b} + (\bvec{v}_E + \bvec{v}_R + \bvec{v}_{\nabla B}),
\end{equation}
%\noindent where $v_{gc,\parallel}$ is the velocity component along the magnetic field direction, $\hat \bvec{b}\equiv \bvec{B}/B$. In an infinitesimal time $\Delta t$, the work $\Delta W$ done on the particle by the electric field is given by $\Delta W = {\Delta t q_e ( \bvec{v}_{gc} \cdot \bvec{E} ) }$, which from combining Eqs.~(\ref{E_ambi_eq})-(\ref{v_gc_eq}) simplifies to the expression:
\noindent where $v_{\parallel}$ is the velocity component along the magnetic field direction, $\hat \bvec{b}\equiv \bvec{B}/B$.
The remaining terms represent the field-perpendicular drifts; i.e. $\bvec{v}_E$, $\bvec{v}_R$, and $\bvec{v}_{\nabla B}$ denote the $\bvec{E}\times\bvec{B}$, curvature, and grad-B drifts respectively:

\begin{eqnarray}\label{drift_eq1}
\bvec{v}_E = \frac{\bvec{E} \times \bvec{B}}{B^2}, \\ \label{drift_eq2}
\bvec{v}_{R} = \frac{m_e v_\parallel^2}{q_e B} \frac{\bvec{R}\times \bvec{B}}{R^2 B}, \\ \label{drift_eq3}
\bvec{v}_{\nabla B} = \frac{m_e v_\perp^2}{2 q_e B} \frac{\bvec{B} \times \nabla B}{B^2}.
\end{eqnarray}

\noindent In the equations above we introduce the electron mass $m_e$, the radius of curvature of a field line $\bvec{R}$, and the parallel and perpendicular velocity components of a particle $v_\parallel$, $v_\perp$. Let us estimate the drifts (Eqs.~\ref{drift_eq1}-\ref{drift_eq3}) for a representative $\sim$300eV particle with pitch angle $\theta $$\sim$$10^\circ$, i.e. with velocity components $v_\parallel$$=$$10^7$m/sec and $v_\perp$$=$$2$$\times$$10^6$m/sec. 
Assuming typical values of plasma parameters found in the magnetosheath (Table~\ref{table_drifts}), we estimate the perpendicular drift speeds to be $|\bvec{v}_E|$$\sim$$10^5$~m/sec, $|\bvec{v}_R|$$\sim$$10^4$~m/sec, $|\bvec{v}_{\nabla B}|$$\sim$$10^3$~m/sec.
%Assuming the fiducial values $|\bvec{E}|$$\sim$$|\bvec{E}_C|$$=$$10^{-3}$V/m, $|\bvec{B}|$$=$$10^{-8}$T,  $|\bvec{R}|$$=$$R_m$$\sim$4$\times$$10^6$m, ${|\bvec{B} \times\nabla B/ B^2|$$=$$10^{-6}$m$^{-1}}$, we estimate the perpendicular drift velocities to be $\bvec{v}_E$$\sim$$10^5$~m/sec, $\bvec{v}_R$$\sim$$10^4$~m/sec, $\bvec{v}_{\nabla B}$$\sim$$10^3$~m/sec.
%We note electric field is dominated by the convective component $\bvec{E}$$\sim$$\bvec{E}_c$, 

\begin{table}
\begin{tabular}{|c|c|}
\hline
 $v_\parallel$ & $10^7$ m/sec  \\ 
 $v_\perp$ & $2\times10^6$ m/sec  \\ 
 $|\bvec{v}_b|$ & $10^5$ m/sec  \\ 
 $|\bvec{E}_C|$ & $10^{-3}$V/m  \\  
 $|\bvec{E}_A|$ & $10^{-5}$V/m  \\  
 $|\bvec{B}|$ & $10^{-8}$T \\    
 $|(\bvec{B}\times \nabla B)/B^2|$ & $10^{-6}$m$^{-1}$ \\    
 $|\bvec{R}|$ & $4\times10^6$m   \\ 
\hline
\end{tabular}
\caption{Typical plasma parameters in the sheath and the velocity of components ($v_\parallel$, $v_\perp$) of a representative $\sim$300eV particle as may be observed by MAVEN's SWEA instrument. These values are used to estimate the magnitude of the perpendicular drifts in sections \ref{discussion_pot_sec}, and may be applied to reduce Eq.~(\ref{W_exact_eq}) to the form (\ref{v_gc_en_eq}) by neglecting small terms.}
\label{table_drifts}
\end{table}

In an infinitesimal time $\Delta t$, the work $\Delta W$ done on the particle by the electric field is given by:

\begin{equation}\label{W_def_eq}
\Delta W = \Delta t q_e ( \bvec{v}_{gc} \cdot \bvec{E} ) ,
\end{equation}

\noindent which from substitution of Eqs.~(\ref{E_ambi_eq}) and (\ref{v_gc_eq}) evaluates to the expression:

\begin{equation}\label{W_exact_eq}
%\Delta W = \Delta t q_e \Big\{ \bvec{v}_{gc} \cdot \bvec{E}_a +(\bvec{v}_E + \bvec{v}_R  + \bvec{v}_{\nabla B} )\cdot \bvec{E}_C  \Big\}
\Delta W = \Delta t q_e \Big\{ \bvec{v}_{gc} \cdot \bvec{E}_A +\Big[\frac{\bvec{E}_A \times \bvec{B}}{B^2} + \bvec{v}_R  + \bvec{v}_{\nabla B} \Big]\cdot \bvec{E}_C  \Big\}
\end{equation}

\noindent \red{As may be estimated from the representative sheath parameters (Table~\ref{table_drifts}), the dominant term of Eq.~(\ref{W_exact_eq}) is the work done by the ambipolar electric field $\bvec{E}_A$. So, we may approximate:}

\begin{equation}\label{v_gc_en_eq}
%\Delta W = \Delta t \frac{v_\parallel}{n_e} (\hat \bvec{b} \cdot \nabla P).
%\Delta W &= \Delta t q_e (v_{\parallel} \hat \bvec{b} \cdot \bvec{E}_a) \\
%\Delta W &\approx \Delta t q_e \{v_{\parallel} \hat \bvec{b} \cdot \bvec{E}_a +(\bvec{v}_{\nabla B} + \bvec{v}_R)\cdot \bvec{E}_C \}
\Delta W \approx \Delta t q_e ( \bvec{v}_{gc} \cdot \bvec{E}_A). 
\end{equation}

%cite lathys paper

%\noindent In Eq.~(\ref{v_gc_en_eq}) we take the perpendicular drifts to be small, allowing the approximation ${\bvec{v}_{gc} \approx v_{\parallel} \hat \bvec{b}}$.
\noindent Comparison of Eqs.(\ref{W_def_eq})-(\ref{v_gc_en_eq}) reveals that electrons primarily ``see'' the ambipolar component of the electric field, $\bvec{E}_A$. 
%As the electron motion is dominated by the parallel component, $v_{gc}$$\sim$$v_\parallel$, 
\red{The systematic} energization comes from the parallel component of the ambipolar field $E_{a,\parallel}$.  As expressed previously, the observation $\Delta\Phi$$\sim$0 then requires some explanation for how electrons moving oppositely along the same field line may be energized by the same amount. If we assume that the ambipolar electric field can be expressed as the gradient of a potential, i.e.,
%Comparison of Eqs.(\ref{W_def_eq})-(\ref{v_gc_en_eq}) reveals that the ambipolar field $\bvec{E}_a(\bvec{x})$ is the only component of $\bvec{E}(\bvec{x})$ in Eq.~(\ref{E_ambi_eq}) that need be considered to explain the particle energization. If we assume that the ambipolar electric field can be expressed as the gradient of a potential, i.e.,
%For the energetic sheath electrons considered here we may take the perpendicular drifts to be small and approximate $\bvec{v}_{gc} \approx v_\parallel \hat \bvec{b}$. Under this approximation, comparison of Eqs.(\ref{W_def_eq})-(\ref{v_gc_en_eq}) reveals that the ambipolar field $\bvec{E}_a(\bvec{x})$ is the only component of $\bvec{E}(\bvec{x})$ in Eq.~(\ref{E_ambi_eq}) that need be considered to explain the particle energization. If we assume that the ambipolar electric field can be expressed as the gradient of a potential, i.e.,

\begin{equation}\label{E_ambi_pot_eq}
\bvec{E}_A(\bvec{x}) = -\nabla \phi,
\end{equation}

\noindent then we need look no further---we have identified a potential field $\phi(\bvec{x})$ capable of energizing the electrons isotropically, in the manner of Eq.~(\ref{phi_electrostatic_eq}). 

The assumed form (\ref{E_ambi_pot_eq}) is not at all far-fetched. Taking the curl of Eq.~\ref{E_ambi_def}, we note the ambipolar electric field will be a potential field (${\nabla\times\bvec{E}_A =\bvec{0}}$) if and only if:

\begin{equation}\label{E_ambi_pot_condition_eq}
\nabla n \times \nabla T_e=0.
\end{equation}

\noindent  This condition (Eq.~\ref{E_ambi_pot_condition_eq}) is quite reasonable, as observations show the variation of $n$ and $T_e$ to be correlated---both quantities exhibit local maxima in the sheath near the $x_{mse}$-axis, and the contours of these quantities will be roughly symmetric about this axis. \comment{citation} We note that if the electron temperature is a function of the density, i.e. if $T_e=T_e(n)$, then the condition (\ref{E_ambi_pot_condition_eq}) will be trivially satisfied. As a special case, $\bvec{E}_A$ will be a potential field if the electrons obey a polytropic equation of state (in which case $T_e(n)$ is a power law). This idea has some precedent, as polytropic models have been applied in Earth's magnetosheath, and the polytropic index has been measured in that system for both ions and electrons \cite{hau93, pang16}. The condition (\ref{E_ambi_pot_condition_eq}) also has the appealing property of preserving the frozen-in flux condition (assumed to apply in this study) even for a non-ideal electric field of the form~(\ref{E_ambi_eq})---see e.g. \citeA{scudder15}.

As mentioned in Section~\ref{obs_sec}, a small systematic correlation between $y_{mse}$ and $\Delta \Phi$ can be observed in Fig.~\ref{delta_phi_4plots_fig}, so that $|\Delta \Phi|$ can be as large as 10-20eV on the flanks. This trend might be accounted for if the ambipolar field is not exactly potential. Alternatively, it may owe to the drifting of electrons through the strong ($|E_C|$$\sim$10$^{-3}$V/m) convective field. 
%Although the ${\bvec{E}\times\bvec{B}}$ drift is not expected to lead to particle energization (see section~\ref{discussion_pot_sec} below), the electrons also experience curvature and gradient drifts.
As estimated above, in the $\bvec{z}_{mse}$=0 plane the curvature ($\bvec{v}_R$) drift may amount to velocities $10^4$m/sec oriented in the +$z_{mse}$ direction, i.e. opposite the convective electric field. We estimate these drifts would cause typical electrons to gain about $\sim$10eV during their entire traversal of the sheath.
%As estimated above, in the $\bvec{z}_{mse}$=0 plane the curvature ($\bvec{v}_R$) and grad-B ($\bvec{v}_{\nabla B}$) drifts combined may amount to velocities $10^4$m/sec oriented in the +$z_{mse}$ direction, i.e. opposite the convective electric field. We estimate these drifts would cause typical electrons to gain about $\sim$10eV during their entire traversal of the sheath.
This effect may be responsible for the slight systematic correlation between $y_{mse}$ and $\Delta \Phi$ observed in Fig.~\ref{delta_phi_4plots_fig}. The mechanism may be roughly imagined as follows: particles with $v_\parallel$$>$0 at a given location in the region $y_{mse}$$>0$ will have spent more time traveling along the field line than particles at the same location with $v_\parallel$$<0$ (which have crossed the bow shock more recently). So, the parallel-propagating electrons will generally have gained more energy via drifting than the anti-parallel electrons in the region $y_{mse}$$>$0, i.e. $\Delta \Phi$$>$0. Similar reasoning may be applied to argue $\Delta \Phi$ should be slightly negative in the region $y_{mse}$$<$0.  We note that the energy gained via the curvature drift is velocity-dependent because of the quadratic dependence of $\bvec{v}_R$ on $v_\parallel$ (Eq.~\ref{drift_eq1}). This could lead to minor departures from our approximation that all electrons moving with a particular orientation with respect to the magnetic field will gain a constant amount of energy (Eq.~).  Detailed investigation of the curvature and gradient drifts, which could contribute a small but non-zero $\Delta \Phi$, is left to future research.

\subsection{Current Feedback}
In another scenario, the observation $\Delta$$\Phi$$\sim$0 might be explained by applying a more self-consistent model of the shock potential. That is, our model of the cross-shock potential $\Phi_s$ (see supplementary document) may not represent a true steady-state, despite being empirically based. Notably, if electrons are energized by different amounts at the two points where the field line meets the shock front, the expected asymmetry of the distribution functions may form a field-parallel current. These currents may lead to a local build-up of charge, and the resulting electric fields would suppress the currents themselves and alter the imposed form of the electric field. Analysis of electron motion in such a self-consistent field might better agree with the observations of $\Delta \Phi$ presented here.
%. Our framework does not include a feedback mechanism that would suppress these currents, and therefore 

% So, the predicted anisotropy presented in Fig.~\ref{phi_anisotropy_model_fig} may not reflect a realistic steady-state even on its own terms.

In order to model the cross-shock potential $\Phi_s$ and our assumed boundary conditions for $f(\bvec{v})$ more realistically, an approach similar to that of \citeA{mitchell14} may be required. In that study, which was concerned with Earth's magnetosheath, the electrons were assumed to be energized entirely by a cross-shock potential $\Delta \Phi^H$. In this kinetic model, the magnitude of $\Delta \Phi^H$ was set throughout the shock to a value that would 1) suppress the parallel current $J_\parallel$ and 2) also satisfy the density predicted by the one-fluid Rankine-Hugoniot relations.  
% ... \comment{(...explain mitchell's method a bit) }. 
Determining whether such a study could reproduce the observed energization and isotropy of the Martian magnetosheath is beyond the scope of this paper. However, we do note that the distributions reported in \citeA{mitchell14} were not highly asymmetric, which is qualitatively consistent with the observations reported here.

Along these lines, we note a promising result from \citeA{mitchell14}: the authors found that the cross-shock potential at Earth on the flanks should not asymptote to zero at infinity, but rather to some constant value. Such a profile at Mars would flatten the potential variation along the shock front, so that the two ends of a given field line would tend to be at more similar potentials (leading to smaller $|\Delta \Phi|$). However, we also note that in their study of Earth's bow shock the authors found that the magnetosheath should settle into an isothermal state; the significant  spatial temperature observed in Mars's magnetosheath\comment{(estimate the total variation, cite key parameters??)} does not agree with this picture.

\section{Summary}\label{summary_sec}

%Earth vs mars
%collisionless? erosion and Steve's ideas...
%results
%pros/cons of diff scenarios

In this paper we analyzed the energization of electrons in the Martian magnetosheath.  The $>$30eV electrons considered in this study move quickly enough to traverse the magnetosheath in about 1~second, so that during this time the field line along which an electron moves is essentially fixed. Due to the different cross-shock potentials at the two locations where the field line intersects the shock \red{(under typical solar wind conditions)}, we may expect to see a significant difference (as much as $\sim$60 eV) between the derived quantities $\Phi_\parallel$ and $\Phi_\downarrow$. The absence of such a signature, as demonstrated for a single orbit (Fig.~\ref{phi_anisotropy_1orb_fig}) and in a statistical average of the $z_{mse}$=0 plane (Fig.~\ref{delta_phi_4plots_fig}), indicates that our basic model of the Martian bow shock needs to be reconsidered.

We presented two possible resolutions for the discrepancy between our model and the observations of $\Delta \Phi$. In one case, we suggested that an ambipolar, (nearly) potential electric field distributed throughout the magnetosheath region could plausibly explain the observation $\Delta \Phi$$\sim$0. In another case, we considered that our predictions for $\Delta \Phi$ would change (and possibly agree better with the observations) if a more self-consistent model for the cross-shock potential $\Phi_s$ were applied. 
Further investigation of these two explanations is beyond the scope of the present paper, which is observational in its focus.
But in any case, we may conclude that diffusive effects such as collisions and wave-particle interactions have a negligible effect on the electron distributions through most of the magnetosheath.  \red{This is based on} the effectiveness of the Liouville mapping technique.

The study was motivated by the simple collisionless model of the sheath developed in \citeA{schwartz19}, which sought to explain the so-called ``erosion'' of the electron flux observed in the inner magnetosheath.
\red{This study suggests that additional acceleration inside the sheath is taking place, obscuring the observational signature that would otherwise be seen if electrons were solely energized at the shock.}
%The present study is founded on similar assumptions, but new predictions for the quantity $\Delta \Phi$ arose by allowing for an arbitrary cone angle of the incoming solar wind.
We are not concerned with the electron flux erosion here. However, we note that if a significant electrostatic field is present in the magnetosheath (as suggested above), incorporating this field's effect may improve the \citeA{schwartz19} model. Not incorporated into this study is the interaction of electrons with neutral hydrogen in the Martian foreshock \cite{mazelle18}.
% As the collisions of electrons with neutrals may perturb the electron distributions in the foreshock, accounting for this effect could potentially improve the quality of future observational analyses.

The significant energization of electrons observed at a planetary bow shock is not unique to the Martian system. 
We speculate that techniques similar to those employed here may be applicable to magnetosheaths at Venus and Earth, for instance.
% We note the approximation used here, that energetic electrons traverse the sheath along a fixed magnetic field, may apply equally well for other planetary magnetosheaths. That is, the distance that a given flux tube moves during one electron traversal---relative to the scale of the magnetosheath---should be insensitive to the scale itself (for a given solar wind speed and electron energy).
No two systems are identical, however, and we foresee that the conditions at other planets may contradict some assumptions applied here. For instance, at Earth one cannot assume that electrons flow along essentially fixed field lines due to the larger shock scale \cite{mitchell12}.
Though such details may complicate the observational analysis, it is nonetheless clear that Liouville mapping can be an effective technique for probing the electric field structure in planetary shocks and magnetosheaths elsewhere in the solar system.

\acknowledgments
%Enter acknowledgments, including your data availability statement, here.
This work was supported by project funds from the NASA MAVEN mission. The solar wind speed, magnetic field, and MAVEN emphemeris were obtained from the MAVEN ``key parameter'' summary data available from the CDAWeb database at https://cdaweb.gsfc.nasa.gov/index.html/. The SWEA pitch angle distributions are available online via the MAVEN Science Data Center at https://lasp.colorado.edu/maven/sdc/public/. 

%% ------------------------------------------------------------------------ %%
%% References and Citations

%%%%%%%%%%%%%%%%%%%%%%%%%%%%%%%%%%%%%%%%%%%%%%%
%
% \bibliography{<name of your .bib file>} don't specify the file extension
%
% don't specify bibliographystyle
%%%%%%%%%%%%%%%%%%%%%%%%%%%%%%%%%%%%%%%%%%%%%%%

\bibliography{maven}

\begin{thebibliography}{}

\bibitem [\protect \citeauthoryear {%
{Acuna}%
\ \protect \BOthers {.}}{%
{Acuna}%
\ \protect \BOthers {.}}{%
{\protect \APACyear {1998}}%
}]{%
acuna98}
\APACinsertmetastar {%
acuna98}%
\begin{APACrefauthors}%
{Acuna}, M\BPBI H.%
, {Connerney}, J\BPBI E\BPBI P.%
, {Wasilewski}, P.%
, {Lin}, R\BPBI P.%
, {Anderson}, K\BPBI A.%
, {Carlson}, C\BPBI W.%
\BDBL {}{Ness}, N\BPBI F.%
\end{APACrefauthors}%
\unskip\
\newblock
\APACrefYearMonthDay{1998}{{\APACmonth{03}}}{}.
\newblock
{\BBOQ}\APACrefatitle {{Magnetic Field and Plasma Observations at Mars: Initial
  Results of}} {{Magnetic Field and Plasma Observations at Mars: Initial
  Results of}}.{\BBCQ}
\newblock
\APACjournalVolNumPages{Science}{279}{}{1676}.
\newblock
\begin{APACrefDOI} \doi{10.1126/science.279.5357.1676} \end{APACrefDOI}
\PrintBackRefs{\CurrentBib}

\bibitem [\protect \citeauthoryear {%
{Connerney}%
\ \protect \BOthers {.}}{%
{Connerney}%
\ \protect \BOthers {.}}{%
{\protect \APACyear {2015}}%
}]{%
connerney15}
\APACinsertmetastar {%
connerney15}%
\begin{APACrefauthors}%
{Connerney}, J\BPBI E\BPBI P.%
, {Espley}, J.%
, {Lawton}, P.%
, {Murphy}, S.%
, {Odom}, J.%
, {Oliversen}, R.%
\BCBL {}\ \BBA {} {Sheppard}, D.%
\end{APACrefauthors}%
\unskip\
\newblock
\APACrefYearMonthDay{2015}{{\APACmonth{12}}}{}.
\newblock
{\BBOQ}\APACrefatitle {{The MAVEN Magnetic Field Investigation}} {{The MAVEN
  Magnetic Field Investigation}}.{\BBCQ}
\newblock
\APACjournalVolNumPages{\ssr}{195}{1-4}{257-291}.
\newblock
\begin{APACrefDOI} \doi{10.1007/s11214-015-0169-4} \end{APACrefDOI}
\PrintBackRefs{\CurrentBib}

\bibitem [\protect \citeauthoryear {%
{Crider}%
\ \protect \BOthers {.}}{%
{Crider}%
\ \protect \BOthers {.}}{%
{\protect \APACyear {2000}}%
}]{%
crider00}
\APACinsertmetastar {%
crider00}%
\begin{APACrefauthors}%
{Crider}, D.%
, {Cloutier}, P.%
, {Law}, C.%
, {Walker}, P.%
, {Chen}, Y.%
, {Acu{\~n}a}, M.%
\BDBL {}{Ness}, N.%
\end{APACrefauthors}%
\unskip\
\newblock
\APACrefYearMonthDay{2000}{{\APACmonth{01}}}{}.
\newblock
{\BBOQ}\APACrefatitle {{Evidence of electron impact ionization in the magnetic
  pileup boundary of Mars}} {{Evidence of electron impact ionization in the
  magnetic pileup boundary of Mars}}.{\BBCQ}
\newblock
\APACjournalVolNumPages{\grl}{27}{1}{45-48}.
\newblock
\begin{APACrefDOI} \doi{10.1029/1999GL003625} \end{APACrefDOI}
\PrintBackRefs{\CurrentBib}

\bibitem [\protect \citeauthoryear {%
{Feldman}%
, {Anderson}%
, {Bame}%
, {Gary}%
\BCBL {}\ \protect \BOthers {.}}{%
{Feldman}%
, {Anderson}%
, {Bame}%
, {Gary}%
\BCBL {}\ \protect \BOthers {.}}{%
{\protect \APACyear {1983}}%
}]{%
feldman83a}
\APACinsertmetastar {%
feldman83a}%
\begin{APACrefauthors}%
{Feldman}, W\BPBI C.%
, {Anderson}, R\BPBI C.%
, {Bame}, S\BPBI J.%
, {Gary}, S\BPBI P.%
, {Gosling}, J\BPBI T.%
, {McComas}, D\BPBI J.%
\BDBL {}{Hoppe}, M\BPBI M.%
\end{APACrefauthors}%
\unskip\
\newblock
\APACrefYearMonthDay{1983}{{\APACmonth{01}}}{}.
\newblock
{\BBOQ}\APACrefatitle {{Electron velocity distributions near the earth's bow
  shock}} {{Electron velocity distributions near the earth's bow
  shock}}.{\BBCQ}
\newblock
\APACjournalVolNumPages{\jgr}{88}{A1}{96-110}.
\newblock
\begin{APACrefDOI} \doi{10.1029/JA088iA01p00096} \end{APACrefDOI}
\PrintBackRefs{\CurrentBib}

\bibitem [\protect \citeauthoryear {%
{Feldman}%
, {Anderson}%
, {Bame}%
, {Gosling}%
\BCBL {}\ \protect \BOthers {.}}{%
{Feldman}%
, {Anderson}%
, {Bame}%
, {Gosling}%
\BCBL {}\ \protect \BOthers {.}}{%
{\protect \APACyear {1983}}%
}]{%
feldman83b}
\APACinsertmetastar {%
feldman83b}%
\begin{APACrefauthors}%
{Feldman}, W\BPBI C.%
, {Anderson}, R\BPBI C.%
, {Bame}, S\BPBI J.%
, {Gosling}, J\BPBI T.%
, {Zwickl}, R\BPBI D.%
\BCBL {}\ \BBA {} {Smith}, E\BPBI J.%
\end{APACrefauthors}%
\unskip\
\newblock
\APACrefYearMonthDay{1983}{{\APACmonth{12}}}{}.
\newblock
{\BBOQ}\APACrefatitle {{Electron velocity distributions near interplanetary
  shocks}} {{Electron velocity distributions near interplanetary
  shocks}}.{\BBCQ}
\newblock
\APACjournalVolNumPages{\jgr}{88}{A12}{9949-9958}.
\newblock
\begin{APACrefDOI} \doi{10.1029/JA088iA12p09949} \end{APACrefDOI}
\PrintBackRefs{\CurrentBib}

\bibitem [\protect \citeauthoryear {%
{Galeev}%
}{%
{Galeev}%
}{%
{\protect \APACyear {1976}}%
}]{%
galeev76}
\APACinsertmetastar {%
galeev76}%
\begin{APACrefauthors}%
{Galeev}, A\BPBI A.%
\end{APACrefauthors}%
\unskip\
\newblock
\APACrefYearMonthDay{1976}{{\APACmonth{01}}}{}.
\newblock
{\BBOQ}\APACrefatitle {{Collisionless shocks.}} {{Collisionless
  shocks.}}{\BBCQ}
\newblock
\BIn{} D\BPBI J.~{Williams}\ (\BED), \APACrefbtitle {Physics of Solar Planetary
  Environments} {Physics of solar planetary environments}\ (\BVOL~1,
  \BPG~464-490).
\PrintBackRefs{\CurrentBib}

\bibitem [\protect \citeauthoryear {%
Halekas%
, Ruhunusiri%
, McFadden%
, Espley%
\BCBL {}\ \BBA {} DiBraccio%
}{%
Halekas%
\ \protect \BOthers {.}}{%
{\protect \APACyear {2019}}%
}]{%
halekas19}
\APACinsertmetastar {%
halekas19}%
\begin{APACrefauthors}%
Halekas, J\BPBI S.%
, Ruhunusiri, S.%
, McFadden, J\BPBI P.%
, Espley, J\BPBI R.%
\BCBL {}\ \BBA {} DiBraccio, G\BPBI A.%
\end{APACrefauthors}%
\unskip\
\newblock
\APACrefYearMonthDay{2019}{}{}.
\newblock
{\BBOQ}\APACrefatitle {Ion Composition Boundary Layer Instabilities at Mars}
  {Ion composition boundary layer instabilities at mars}.{\BBCQ}
\newblock
\APACjournalVolNumPages{Geophysical Research Letters}{46}{17-18}{10303-10312}.
\newblock
\begin{APACrefURL}
  \url{https://agupubs.onlinelibrary.wiley.com/doi/abs/10.1029/2019GL084779}
  \end{APACrefURL}
\newblock
\begin{APACrefDOI} \doi{10.1029/2019GL084779} \end{APACrefDOI}
\PrintBackRefs{\CurrentBib}

\bibitem [\protect \citeauthoryear {%
{Halekas}%
\ \protect \BOthers {.}}{%
{Halekas}%
\ \protect \BOthers {.}}{%
{\protect \APACyear {2015}}%
}]{%
halekas15}
\APACinsertmetastar {%
halekas15}%
\begin{APACrefauthors}%
{Halekas}, J\BPBI S.%
, {Taylor}, E\BPBI R.%
, {Dalton}, G.%
, {Johnson}, G.%
, {Curtis}, D\BPBI W.%
, {McFadden}, J\BPBI P.%
\BDBL {}{Jakosky}, B\BPBI M.%
\end{APACrefauthors}%
\unskip\
\newblock
\APACrefYearMonthDay{2015}{{\APACmonth{12}}}{}.
\newblock
{\BBOQ}\APACrefatitle {{The Solar Wind Ion Analyzer for MAVEN}} {{The Solar
  Wind Ion Analyzer for MAVEN}}.{\BBCQ}
\newblock
\APACjournalVolNumPages{\ssr}{195}{1-4}{125-151}.
\newblock
\begin{APACrefDOI} \doi{10.1007/s11214-013-0029-z} \end{APACrefDOI}
\PrintBackRefs{\CurrentBib}

\bibitem [\protect \citeauthoryear {%
{Hau}%
, {Phan}%
, {Sonnerup}%
\BCBL {}\ \BBA {} {Paschmann}%
}{%
{Hau}%
\ \protect \BOthers {.}}{%
{\protect \APACyear {1993}}%
}]{%
hau93}
\APACinsertmetastar {%
hau93}%
\begin{APACrefauthors}%
{Hau}, L\BPBI N.%
, {Phan}, T\BPBI D.%
, {Sonnerup}, B\BPBI U\BPBI O.%
\BCBL {}\ \BBA {} {Paschmann}, G.%
\end{APACrefauthors}%
\unskip\
\newblock
\APACrefYearMonthDay{1993}{{\APACmonth{10}}}{}.
\newblock
{\BBOQ}\APACrefatitle {{Double-polytropic closure in the magnetosheath}}
  {{Double-polytropic closure in the magnetosheath}}.{\BBCQ}
\newblock
\APACjournalVolNumPages{\grl}{20}{20}{2255-2258}.
\newblock
\begin{APACrefDOI} \doi{10.1029/93GL02491} \end{APACrefDOI}
\PrintBackRefs{\CurrentBib}

\bibitem [\protect \citeauthoryear {%
{Jakosky}%
\ \protect \BOthers {.}}{%
{Jakosky}%
\ \protect \BOthers {.}}{%
{\protect \APACyear {2015}}%
}]{%
jakosky15}
\APACinsertmetastar {%
jakosky15}%
\begin{APACrefauthors}%
{Jakosky}, B\BPBI M.%
, {Lin}, R\BPBI P.%
, {Grebowsky}, J\BPBI M.%
, {Luhmann}, J\BPBI G.%
, {Mitchell}, D\BPBI F.%
, {Beutelschies}, G.%
\BDBL {}{Zurek}, R.%
\end{APACrefauthors}%
\unskip\
\newblock
\APACrefYearMonthDay{2015}{{\APACmonth{12}}}{}.
\newblock
{\BBOQ}\APACrefatitle {{The Mars Atmosphere and Volatile Evolution ( MAVEN)
  Mission}} {{The Mars Atmosphere and Volatile Evolution ( MAVEN)
  Mission}}.{\BBCQ}
\newblock
\APACjournalVolNumPages{\ssr}{195}{1-4}{3-48}.
\newblock
\begin{APACrefDOI} \doi{10.1007/s11214-015-0139-x} \end{APACrefDOI}
\PrintBackRefs{\CurrentBib}

\bibitem [\protect \citeauthoryear {%
{Kivelson}%
\ \BBA {} {Russell}%
}{%
{Kivelson}%
\ \BBA {} {Russell}%
}{%
{\protect \APACyear {1995}}%
}]{%
kivelson95}
\APACinsertmetastar {%
kivelson95}%
\begin{APACrefauthors}%
{Kivelson}, M\BPBI G.%
\BCBT {}\ \BBA {} {Russell}, C\BPBI T.%
\end{APACrefauthors}%
\unskip\
\newblock
\APACrefYear{1995}.
\newblock
\APACrefbtitle {{Introduction to Space Physics}} {{Introduction to Space
  Physics}}.
\PrintBackRefs{\CurrentBib}

\bibitem [\protect \citeauthoryear {%
Koenders%
, Goetz%
, Richter%
, Motschmann%
\BCBL {}\ \BBA {} Glassmeier%
}{%
Koenders%
\ \protect \BOthers {.}}{%
{\protect \APACyear {2016}}%
}]{%
koenders16}
\APACinsertmetastar {%
koenders16}%
\begin{APACrefauthors}%
Koenders, C.%
, Goetz, C.%
, Richter, I.%
, Motschmann, U.%
\BCBL {}\ \BBA {} Glassmeier, K\BHBI H.%
\end{APACrefauthors}%
\unskip\
\newblock
\APACrefYearMonthDay{2016}{10}{}.
\newblock
{\BBOQ}\APACrefatitle {{Magnetic field pile-up and draping at intermediately
  active comets: results from comet 67P/Churyumov–Gerasimenko at 2.0 AU}}
  {{Magnetic field pile-up and draping at intermediately active comets: results
  from comet 67P/Churyumov–Gerasimenko at 2.0 AU}}.{\BBCQ}
\newblock
\APACjournalVolNumPages{Monthly Notices of the Royal Astronomical
  Society}{462}{Suppl\_1}{S235-S241}.
\newblock
\begin{APACrefURL} \url{https://doi.org/10.1093/mnras/stw2480} \end{APACrefURL}
\newblock
\begin{APACrefDOI} \doi{10.1093/mnras/stw2480} \end{APACrefDOI}
\PrintBackRefs{\CurrentBib}

\bibitem [\protect \citeauthoryear {%
{Lefebvre}%
, {Schwartz}%
, {Fazakerley}%
\BCBL {}\ \BBA {} {D{\'e}cr{\'e}au}%
}{%
{Lefebvre}%
\ \protect \BOthers {.}}{%
{\protect \APACyear {2007}}%
}]{%
lefebvre07}
\APACinsertmetastar {%
lefebvre07}%
\begin{APACrefauthors}%
{Lefebvre}, B.%
, {Schwartz}, S\BPBI J.%
, {Fazakerley}, A\BPBI F.%
\BCBL {}\ \BBA {} {D{\'e}cr{\'e}au}, P.%
\end{APACrefauthors}%
\unskip\
\newblock
\APACrefYearMonthDay{2007}{{\APACmonth{09}}}{}.
\newblock
{\BBOQ}\APACrefatitle {{Electron dynamics and cross-shock potential at the
  quasi-perpendicular Earth's bow shock}} {{Electron dynamics and cross-shock
  potential at the quasi-perpendicular Earth's bow shock}}.{\BBCQ}
\newblock
\APACjournalVolNumPages{Journal of Geophysical Research (Space
  Physics)}{112}{A9}{A09212}.
\newblock
\begin{APACrefDOI} \doi{10.1029/2007JA012277} \end{APACrefDOI}
\PrintBackRefs{\CurrentBib}

\bibitem [\protect \citeauthoryear {%
Lundin%
\ \protect \BOthers {.}}{%
Lundin%
\ \protect \BOthers {.}}{%
{\protect \APACyear {2004}}%
}]{%
lundin04}
\APACinsertmetastar {%
lundin04}%
\begin{APACrefauthors}%
Lundin, R.%
, Barabash, S.%
, Andersson, H.%
, Holmstr{\"o}m, M.%
, Grigoriev, A.%
, Yamauchi, M.%
\BDBL {}others%
\end{APACrefauthors}%
\unskip\
\newblock
\APACrefYearMonthDay{2004}{}{}.
\newblock
{\BBOQ}\APACrefatitle {Solar wind-induced atmospheric erosion at Mars: First
  results from ASPERA-3 on Mars Express} {Solar wind-induced atmospheric
  erosion at mars: First results from aspera-3 on mars express}.{\BBCQ}
\newblock
\APACjournalVolNumPages{science}{305}{5692}{1933--1936}.
\PrintBackRefs{\CurrentBib}

\bibitem [\protect \citeauthoryear {%
C.~{Mazelle}%
\ \protect \BOthers {.}}{%
C.~{Mazelle}%
\ \protect \BOthers {.}}{%
{\protect \APACyear {2004}}%
}]{%
mazelle04}
\APACinsertmetastar {%
mazelle04}%
\begin{APACrefauthors}%
{Mazelle}, C.%
, {Winterhalter}, D.%
, {Sauer}, K.%
, {Trotignon}, J\BPBI G.%
, {Acu{\~n}a}, M\BPBI H.%
, {Baumg{\"a}rtel}, K.%
\BDBL {}{Slavin}, J.%
\end{APACrefauthors}%
\unskip\
\newblock
\APACrefYearMonthDay{2004}{{\APACmonth{03}}}{}.
\newblock
{\BBOQ}\APACrefatitle {{Bow Shock and Upstream Phenomena at Mars}} {{Bow Shock
  and Upstream Phenomena at Mars}}.{\BBCQ}
\newblock
\APACjournalVolNumPages{\ssr}{111}{1}{115-181}.
\newblock
\begin{APACrefDOI} \doi{10.1023/B:SPAC.0000032717.98679.d0} \end{APACrefDOI}
\PrintBackRefs{\CurrentBib}

\bibitem [\protect \citeauthoryear {%
C\BPBI X.~{Mazelle}%
\ \protect \BOthers {.}}{%
C\BPBI X.~{Mazelle}%
\ \protect \BOthers {.}}{%
{\protect \APACyear {2018}}%
}]{%
mazelle18}
\APACinsertmetastar {%
mazelle18}%
\begin{APACrefauthors}%
{Mazelle}, C\BPBI X.%
, {Meziane}, K.%
, {Mitchell}, D\BPBI L.%
, {Garnier}, P.%
, {Espley}, J\BPBI R.%
, {Hamza}, A\BPBI M.%
\BDBL {}{Jakosky}, B\BPBI M.%
\end{APACrefauthors}%
\unskip\
\newblock
\APACrefYearMonthDay{2018}{{\APACmonth{05}}}{}.
\newblock
{\BBOQ}\APACrefatitle {{Evidence for Neutrals-Foreshock Electrons Impact at
  Mars}} {{Evidence for Neutrals-Foreshock Electrons Impact at Mars}}.{\BBCQ}
\newblock
\APACjournalVolNumPages{\grl}{45}{9}{3768-3774}.
\newblock
\begin{APACrefDOI} \doi{10.1002/2018GL077298} \end{APACrefDOI}
\PrintBackRefs{\CurrentBib}

\bibitem [\protect \citeauthoryear {%
D\BPBI L.~{Mitchell}%
\ \protect \BOthers {.}}{%
D\BPBI L.~{Mitchell}%
\ \protect \BOthers {.}}{%
{\protect \APACyear {2016}}%
}]{%
mitchell16}
\APACinsertmetastar {%
mitchell16}%
\begin{APACrefauthors}%
{Mitchell}, D\BPBI L.%
, {Mazelle}, C.%
, {Sauvaud}, J\BPBI A.%
, {Thocaven}, J\BPBI J.%
, {Rouzaud}, J.%
, {Fedorov}, A.%
\BDBL {}{Jakosky}, B\BPBI M.%
\end{APACrefauthors}%
\unskip\
\newblock
\APACrefYearMonthDay{2016}{{\APACmonth{04}}}{}.
\newblock
{\BBOQ}\APACrefatitle {{The MAVEN Solar Wind Electron Analyzer}} {{The MAVEN
  Solar Wind Electron Analyzer}}.{\BBCQ}
\newblock
\APACjournalVolNumPages{\ssr}{200}{1-4}{495-528}.
\newblock
\begin{APACrefDOI} \doi{10.1007/s11214-015-0232-1} \end{APACrefDOI}
\PrintBackRefs{\CurrentBib}

\bibitem [\protect \citeauthoryear {%
J\BPBI J.~{Mitchell}%
\ \BBA {} {Schwartz}%
}{%
J\BPBI J.~{Mitchell}%
\ \BBA {} {Schwartz}%
}{%
{\protect \APACyear {2013}}%
}]{%
mitchell13}
\APACinsertmetastar {%
mitchell13}%
\begin{APACrefauthors}%
{Mitchell}, J\BPBI J.%
\BCBT {}\ \BBA {} {Schwartz}, S\BPBI J.%
\end{APACrefauthors}%
\unskip\
\newblock
\APACrefYearMonthDay{2013}{{\APACmonth{12}}}{}.
\newblock
{\BBOQ}\APACrefatitle {{Nonlocal electron heating at the Earth's bow shock and
  the role of the magnetically tangent point}} {{Nonlocal electron heating at
  the Earth's bow shock and the role of the magnetically tangent
  point}}.{\BBCQ}
\newblock
\APACjournalVolNumPages{Journal of Geophysical Research (Space
  Physics)}{118}{12}{7566-7575}.
\newblock
\begin{APACrefDOI} \doi{10.1002/2013JA019226} \end{APACrefDOI}
\PrintBackRefs{\CurrentBib}

\bibitem [\protect \citeauthoryear {%
J\BPBI J.~{Mitchell}%
\ \BBA {} {Schwartz}%
}{%
J\BPBI J.~{Mitchell}%
\ \BBA {} {Schwartz}%
}{%
{\protect \APACyear {2014}}%
}]{%
mitchell14}
\APACinsertmetastar {%
mitchell14}%
\begin{APACrefauthors}%
{Mitchell}, J\BPBI J.%
\BCBT {}\ \BBA {} {Schwartz}, S\BPBI J.%
\end{APACrefauthors}%
\unskip\
\newblock
\APACrefYearMonthDay{2014}{{\APACmonth{02}}}{}.
\newblock
{\BBOQ}\APACrefatitle {{Isothermal magnetosheath electrons due to nonlocal
  electron cross talk}} {{Isothermal magnetosheath electrons due to nonlocal
  electron cross talk}}.{\BBCQ}
\newblock
\APACjournalVolNumPages{Journal of Geophysical Research (Space
  Physics)}{119}{2}{1080-1093}.
\newblock
\begin{APACrefDOI} \doi{10.1002/2013JA019211} \end{APACrefDOI}
\PrintBackRefs{\CurrentBib}

\bibitem [\protect \citeauthoryear {%
J\BPBI J.~{Mitchell}%
, {Schwartz}%
\BCBL {}\ \BBA {} {Auster}%
}{%
J\BPBI J.~{Mitchell}%
\ \protect \BOthers {.}}{%
{\protect \APACyear {2012}}%
}]{%
mitchell12}
\APACinsertmetastar {%
mitchell12}%
\begin{APACrefauthors}%
{Mitchell}, J\BPBI J.%
, {Schwartz}, S\BPBI J.%
\BCBL {}\ \BBA {} {Auster}, U.%
\end{APACrefauthors}%
\unskip\
\newblock
\APACrefYearMonthDay{2012}{{\APACmonth{03}}}{}.
\newblock
{\BBOQ}\APACrefatitle {{Electron cross talk and asymmetric electron
  distributions near the Earth's bowshock}} {{Electron cross talk and
  asymmetric electron distributions near the Earth's bowshock}}.{\BBCQ}
\newblock
\APACjournalVolNumPages{Annales Geophysicae}{30}{3}{503-513}.
\newblock
\begin{APACrefDOI} \doi{10.5194/angeo-30-503-2012} \end{APACrefDOI}
\PrintBackRefs{\CurrentBib}

\bibitem [\protect \citeauthoryear {%
{Montgomery}%
, {Asbridge}%
\BCBL {}\ \BBA {} {Bame}%
}{%
{Montgomery}%
\ \protect \BOthers {.}}{%
{\protect \APACyear {1970}}%
}]{%
montgomery70}
\APACinsertmetastar {%
montgomery70}%
\begin{APACrefauthors}%
{Montgomery}, M\BPBI D.%
, {Asbridge}, J\BPBI R.%
\BCBL {}\ \BBA {} {Bame}, S\BPBI J.%
\end{APACrefauthors}%
\unskip\
\newblock
\APACrefYearMonthDay{1970}{{\APACmonth{01}}}{}.
\newblock
{\BBOQ}\APACrefatitle {{Vela 4 plasma observations near the
  Earth{\textquoteright}s bow shock}} {{Vela 4 plasma observations near the
  Earth{\textquoteright}s bow shock}}.{\BBCQ}
\newblock
\APACjournalVolNumPages{\jgr}{75}{7}{1217}.
\newblock
\begin{APACrefDOI} \doi{10.1029/JA075i007p01217} \end{APACrefDOI}
\PrintBackRefs{\CurrentBib}

\bibitem [\protect \citeauthoryear {%
{Nagy}%
\ \protect \BOthers {.}}{%
{Nagy}%
\ \protect \BOthers {.}}{%
{\protect \APACyear {2004}}%
}]{%
nagy04}
\APACinsertmetastar {%
nagy04}%
\begin{APACrefauthors}%
{Nagy}, A\BPBI F.%
, {Winterhalter}, D.%
, {Sauer}, K.%
, {Cravens}, T\BPBI E.%
, {Brecht}, S.%
, {Mazelle}, C.%
\BDBL {}{Trotignon}, J\BPBI G.%
\end{APACrefauthors}%
\unskip\
\newblock
\APACrefYearMonthDay{2004}{{\APACmonth{03}}}{}.
\newblock
{\BBOQ}\APACrefatitle {{The plasma Environment of Mars}} {{The plasma
  Environment of Mars}}.{\BBCQ}
\newblock
\APACjournalVolNumPages{\ssr}{111}{1}{33-114}.
\newblock
\begin{APACrefDOI} \doi{10.1023/B:SPAC.0000032718.47512.92} \end{APACrefDOI}
\PrintBackRefs{\CurrentBib}

\bibitem [\protect \citeauthoryear {%
{Pang}%
, {Cao}%
\BCBL {}\ \BBA {} {Ma}%
}{%
{Pang}%
\ \protect \BOthers {.}}{%
{\protect \APACyear {2016}}%
}]{%
pang16}
\APACinsertmetastar {%
pang16}%
\begin{APACrefauthors}%
{Pang}, X.%
, {Cao}, J.%
\BCBL {}\ \BBA {} {Ma}, Y.%
\end{APACrefauthors}%
\unskip\
\newblock
\APACrefYearMonthDay{2016}{{\APACmonth{03}}}{}.
\newblock
{\BBOQ}\APACrefatitle {{Polytropic index of magnetosheath ions based on
  homogeneous MHD Bernoulli Integral}} {{Polytropic index of magnetosheath ions
  based on homogeneous MHD Bernoulli Integral}}.{\BBCQ}
\newblock
\APACjournalVolNumPages{Journal of Geophysical Research (Space
  Physics)}{121}{3}{2349-2359}.
\newblock
\begin{APACrefDOI} \doi{10.1002/2015JA022303} \end{APACrefDOI}
\PrintBackRefs{\CurrentBib}

\bibitem [\protect \citeauthoryear {%
{Sagdeev}%
}{%
{Sagdeev}%
}{%
{\protect \APACyear {1966}}%
}]{%
sagdeev66}
\APACinsertmetastar {%
sagdeev66}%
\begin{APACrefauthors}%
{Sagdeev}, R\BPBI Z.%
\end{APACrefauthors}%
\unskip\
\newblock
\APACrefYearMonthDay{1966}{{\APACmonth{01}}}{}.
\newblock
{\BBOQ}\APACrefatitle {{Cooperative Phenomena and Shock Waves in Collisionless
  Plasmas}} {{Cooperative Phenomena and Shock Waves in Collisionless
  Plasmas}}.{\BBCQ}
\newblock
\APACjournalVolNumPages{Reviews of Plasma Physics}{4}{}{23}.
\PrintBackRefs{\CurrentBib}

\bibitem [\protect \citeauthoryear {%
{Schwartz}%
\ \protect \BOthers {.}}{%
{Schwartz}%
\ \protect \BOthers {.}}{%
{\protect \APACyear {2019}}%
}]{%
schwartz19}
\APACinsertmetastar {%
schwartz19}%
\begin{APACrefauthors}%
{Schwartz}, S\BPBI J.%
, {Andersson}, L.%
, {Xu}, S.%
, {Mitchell}, D\BPBI L.%
, {Akbari}, H.%
, {Ergun}, R\BPBI E.%
\BDBL {}{Meziane}, K.%
\end{APACrefauthors}%
\unskip\
\newblock
\APACrefYearMonthDay{2019}{{\APACmonth{11}}}{}.
\newblock
{\BBOQ}\APACrefatitle {{Collisionless Electron Dynamics in the Magnetosheath of
  Mars}} {{Collisionless Electron Dynamics in the Magnetosheath of
  Mars}}.{\BBCQ}
\newblock
\APACjournalVolNumPages{\grl}{46}{21}{11,679-11,688}.
\newblock
\begin{APACrefDOI} \doi{10.1029/2019GL085037} \end{APACrefDOI}
\PrintBackRefs{\CurrentBib}

\bibitem [\protect \citeauthoryear {%
{Schwartz}%
, {Daly}%
\BCBL {}\ \BBA {} {Fazakerley}%
}{%
{Schwartz}%
\ \protect \BOthers {.}}{%
{\protect \APACyear {1998}}%
}]{%
schwartz98}
\APACinsertmetastar {%
schwartz98}%
\begin{APACrefauthors}%
{Schwartz}, S\BPBI J.%
, {Daly}, P\BPBI W.%
\BCBL {}\ \BBA {} {Fazakerley}, A\BPBI N.%
\end{APACrefauthors}%
\unskip\
\newblock
\APACrefYearMonthDay{1998}{{\APACmonth{01}}}{}.
\newblock
{\BBOQ}\APACrefatitle {{Multi-Spacecraft Analysis of Plasma Kinetics}}
  {{Multi-Spacecraft Analysis of Plasma Kinetics}}.{\BBCQ}
\newblock
\APACjournalVolNumPages{ISSI Scientific Reports Series}{1}{}{159-184}.
\PrintBackRefs{\CurrentBib}

\bibitem [\protect \citeauthoryear {%
{Scudder}%
, {Karimabadi}%
, {Daughton}%
\BCBL {}\ \BBA {} {Roytershteyn}%
}{%
{Scudder}%
\ \protect \BOthers {.}}{%
{\protect \APACyear {2015}}%
}]{%
scudder15}
\APACinsertmetastar {%
scudder15}%
\begin{APACrefauthors}%
{Scudder}, J\BPBI D.%
, {Karimabadi}, H.%
, {Daughton}, W.%
\BCBL {}\ \BBA {} {Roytershteyn}, V.%
\end{APACrefauthors}%
\unskip\
\newblock
\APACrefYearMonthDay{2015}{{\APACmonth{10}}}{}.
\newblock
{\BBOQ}\APACrefatitle {{Frozen flux violation, electron demagnetization and
  magnetic reconnection}} {{Frozen flux violation, electron demagnetization and
  magnetic reconnection}}.{\BBCQ}
\newblock
\APACjournalVolNumPages{Physics of Plasmas}{22}{10}{101204}.
\newblock
\begin{APACrefDOI} \doi{10.1063/1.4932332} \end{APACrefDOI}
\PrintBackRefs{\CurrentBib}

\bibitem [\protect \citeauthoryear {%
{Scudder}%
, {Mangeney}%
, {Lacombe}%
, {Harvey}%
\BCBL {}\ \BBA {} {Aggson}%
}{%
{Scudder}%
\ \protect \BOthers {.}}{%
{\protect \APACyear {1986}}%
}]{%
scudder86b}
\APACinsertmetastar {%
scudder86b}%
\begin{APACrefauthors}%
{Scudder}, J\BPBI D.%
, {Mangeney}, A.%
, {Lacombe}, C.%
, {Harvey}, C\BPBI C.%
\BCBL {}\ \BBA {} {Aggson}, T\BPBI L.%
\end{APACrefauthors}%
\unskip\
\newblock
\APACrefYearMonthDay{1986}{{\APACmonth{10}}}{}.
\newblock
{\BBOQ}\APACrefatitle {{The resolved layer of a collisionless, high
  {\ensuremath{\beta}}, supercritical, quasi-perpendicular shock wave. 2.
  Dissipative fluid electrodynamics}} {{The resolved layer of a collisionless,
  high {\ensuremath{\beta}}, supercritical, quasi-perpendicular shock wave. 2.
  Dissipative fluid electrodynamics}}.{\BBCQ}
\newblock
\APACjournalVolNumPages{\jgr}{91}{A10}{11053-11074}.
\newblock
\begin{APACrefDOI} \doi{10.1029/JA091iA10p11053} \end{APACrefDOI}
\PrintBackRefs{\CurrentBib}

\bibitem [\protect \citeauthoryear {%
{Spreiter}%
, {Summers}%
\BCBL {}\ \BBA {} {Alksne}%
}{%
{Spreiter}%
\ \protect \BOthers {.}}{%
{\protect \APACyear {1966}}%
}]{%
spreiter66}
\APACinsertmetastar {%
spreiter66}%
\begin{APACrefauthors}%
{Spreiter}, J\BPBI R.%
, {Summers}, A\BPBI L.%
\BCBL {}\ \BBA {} {Alksne}, A\BPBI Y.%
\end{APACrefauthors}%
\unskip\
\newblock
\APACrefYearMonthDay{1966}{{\APACmonth{03}}}{}.
\newblock
{\BBOQ}\APACrefatitle {{Hydromagnetic flow around the magnetosphere}}
  {{Hydromagnetic flow around the magnetosphere}}.{\BBCQ}
\newblock
\APACjournalVolNumPages{\planss}{14}{3}{223,IN1,251-250,IN2,253}.
\newblock
\begin{APACrefDOI} \doi{10.1016/0032-0633(66)90124-3} \end{APACrefDOI}
\PrintBackRefs{\CurrentBib}

\bibitem [\protect \citeauthoryear {%
{Vignes}%
\ \protect \BOthers {.}}{%
{Vignes}%
\ \protect \BOthers {.}}{%
{\protect \APACyear {2000}}%
}]{%
vignes00}
\APACinsertmetastar {%
vignes00}%
\begin{APACrefauthors}%
{Vignes}, D.%
, {Mazelle}, C.%
, {Reme}, H.%
, {Acu{\~n}a}, M\BPBI H.%
, {Connerney}, J\BPBI E\BPBI P.%
, {Lin}, R\BPBI P.%
\BDBL {}{Ness}, N\BPBI F.%
\end{APACrefauthors}%
\unskip\
\newblock
\APACrefYearMonthDay{2000}{{\APACmonth{01}}}{}.
\newblock
{\BBOQ}\APACrefatitle {{The solar wind interaction with Mars: Locations and
  shapes of the bow shock and the magnetic pile-up boundary from the
  observations of the MAG/ER Experiment onboard Mars Global Surveyor}} {{The
  solar wind interaction with Mars: Locations and shapes of the bow shock and
  the magnetic pile-up boundary from the observations of the MAG/ER Experiment
  onboard Mars Global Surveyor}}.{\BBCQ}
\newblock
\APACjournalVolNumPages{\grl}{27}{1}{49-52}.
\newblock
\begin{APACrefDOI} \doi{10.1029/1999GL010703} \end{APACrefDOI}
\PrintBackRefs{\CurrentBib}

\end{thebibliography}


\begin{thebibliography}{}

\bibitem [\protect \citeauthoryear {%
{Barlow}%
}{%
{Barlow}%
}{%
{\protect \APACyear {1989}}%
}]{%
barlow89}
\APACinsertmetastar {%
barlow89}%
\begin{APACrefauthors}%
{Barlow}, R.%
\end{APACrefauthors}%
\unskip\
\newblock
\APACrefYear{1989}.
\newblock
\APACrefbtitle {{Statistics. A guide to the use of statistical methods in the
  physical sciences}} {{Statistics. A guide to the use of statistical methods
  in the physical sciences}}.
\PrintBackRefs{\CurrentBib}

\bibitem [\protect \citeauthoryear {%
D.~{Crider}%
\ \protect \BOthers {.}}{%
D.~{Crider}%
\ \protect \BOthers {.}}{%
{\protect \APACyear {2000}}%
}]{%
crider00}
\APACinsertmetastar {%
crider00}%
\begin{APACrefauthors}%
{Crider}, D.%
, {Cloutier}, P.%
, {Law}, C.%
, {Walker}, P.%
, {Chen}, Y.%
, {Acu{\~n}a}, M.%
\BDBL {}{Ness}, N.%
\end{APACrefauthors}%
\unskip\
\newblock
\APACrefYearMonthDay{2000}{{\APACmonth{01}}}{}.
\newblock
{\BBOQ}\APACrefatitle {{Evidence of electron impact ionization in the magnetic
  pileup boundary of Mars}} {{Evidence of electron impact ionization in the
  magnetic pileup boundary of Mars}}.{\BBCQ}
\newblock
\APACjournalVolNumPages{\grl}{27}{1}{45-48}.
\newblock
\begin{APACrefDOI} \doi{10.1029/1999GL003625} \end{APACrefDOI}
\PrintBackRefs{\CurrentBib}

\bibitem [\protect \citeauthoryear {%
D\BPBI H.~{Crider}%
\ \protect \BOthers {.}}{%
D\BPBI H.~{Crider}%
\ \protect \BOthers {.}}{%
{\protect \APACyear {2004}}%
}]{%
crider04}
\APACinsertmetastar {%
crider04}%
\begin{APACrefauthors}%
{Crider}, D\BPBI H.%
, {Brain}, D\BPBI A.%
, {Acu{\~n}a}, M\BPBI H.%
, {Vignes}, D.%
, {Mazelle}, C.%
\BCBL {}\ \BBA {} {Bertucci}, C.%
\end{APACrefauthors}%
\unskip\
\newblock
\APACrefYearMonthDay{2004}{{\APACmonth{03}}}{}.
\newblock
{\BBOQ}\APACrefatitle {{Mars Global Surveyor Observations of Solar Wind
  Magnetic Field Draping Around Mars}} {{Mars Global Surveyor Observations of
  Solar Wind Magnetic Field Draping Around Mars}}.{\BBCQ}
\newblock
\APACjournalVolNumPages{\ssr}{111}{1}{203-221}.
\newblock
\begin{APACrefDOI} \doi{10.1023/B:SPAC.0000032714.66124.4e} \end{APACrefDOI}
\PrintBackRefs{\CurrentBib}

\bibitem [\protect \citeauthoryear {%
{Daly}%
, {Schwartz}%
\BCBL {}\ \BBA {} {Lefebvre}%
}{%
{Daly}%
\ \protect \BOthers {.}}{%
{\protect \APACyear {2008}}%
}]{%
daly08}
\APACinsertmetastar {%
daly08}%
\begin{APACrefauthors}%
{Daly}, P\BPBI W.%
, {Schwartz}, S\BPBI J.%
\BCBL {}\ \BBA {} {Lefebvre}, B.%
\end{APACrefauthors}%
\unskip\
\newblock
\APACrefYearMonthDay{2008}{{\APACmonth{01}}}{}.
\newblock
{\BBOQ}\APACrefatitle {{Plasma Kinetics}} {{Plasma Kinetics}}.{\BBCQ}
\newblock
\APACjournalVolNumPages{ISSI Scientific Reports Series}{8}{}{75-80}.
\PrintBackRefs{\CurrentBib}

\bibitem [\protect \citeauthoryear {%
{Feldman}%
\ \protect \BOthers {.}}{%
{Feldman}%
\ \protect \BOthers {.}}{%
{\protect \APACyear {1983}}%
}]{%
feldman83a}
\APACinsertmetastar {%
feldman83a}%
\begin{APACrefauthors}%
{Feldman}, W\BPBI C.%
, {Anderson}, R\BPBI C.%
, {Bame}, S\BPBI J.%
, {Gary}, S\BPBI P.%
, {Gosling}, J\BPBI T.%
, {McComas}, D\BPBI J.%
\BDBL {}{Hoppe}, M\BPBI M.%
\end{APACrefauthors}%
\unskip\
\newblock
\APACrefYearMonthDay{1983}{{\APACmonth{01}}}{}.
\newblock
{\BBOQ}\APACrefatitle {{Electron velocity distributions near the earth's bow
  shock}} {{Electron velocity distributions near the earth's bow
  shock}}.{\BBCQ}
\newblock
\APACjournalVolNumPages{\jgr}{88}{A1}{96-110}.
\newblock
\begin{APACrefDOI} \doi{10.1029/JA088iA01p00096} \end{APACrefDOI}
\PrintBackRefs{\CurrentBib}

\bibitem [\protect \citeauthoryear {%
{Knight}%
}{%
{Knight}%
}{%
{\protect \APACyear {1973}}%
}]{%
knight73}
\APACinsertmetastar {%
knight73}%
\begin{APACrefauthors}%
{Knight}, S.%
\end{APACrefauthors}%
\unskip\
\newblock
\APACrefYearMonthDay{1973}{{\APACmonth{05}}}{}.
\newblock
{\BBOQ}\APACrefatitle {{Parallel electric fields}} {{Parallel electric
  fields}}.{\BBCQ}
\newblock
\APACjournalVolNumPages{\planss}{21}{5}{741-750}.
\newblock
\begin{APACrefDOI} \doi{10.1016/0032-0633(73)90093-7} \end{APACrefDOI}
\PrintBackRefs{\CurrentBib}

\bibitem [\protect \citeauthoryear {%
{Luhmann}%
\ \protect \BOthers {.}}{%
{Luhmann}%
\ \protect \BOthers {.}}{%
{\protect \APACyear {2015}}%
}]{%
luhmann15}
\APACinsertmetastar {%
luhmann15}%
\begin{APACrefauthors}%
{Luhmann}, J\BPBI G.%
, {Dong}, C.%
, {Ma}, Y.%
, {Curry}, S.%
, {Mitchell}, D\BPBI L.%
, {Mazelle}, C\BPBI X.%
\BDBL {}{Jakosky}, B\BPBI M.%
\end{APACrefauthors}%
\unskip\
\newblock
\APACrefYearMonthDay{2015}{{\APACmonth{12}}}{}.
\newblock
{\BBOQ}\APACrefatitle {{Implications of MAVEN Mars Near-Wake Measurements and
  Models}} {{Implications of MAVEN Mars Near-Wake Measurements and
  Models}}.{\BBCQ}
\newblock
\BIn{} \APACrefbtitle {AGU Fall Meeting Abstracts} {Agu fall meeting
  abstracts}\ (\BVOL\ 2015, \BPG~P21A-2073).
\PrintBackRefs{\CurrentBib}

\bibitem [\protect \citeauthoryear {%
{Schwartz}%
\ \protect \BOthers {.}}{%
{Schwartz}%
\ \protect \BOthers {.}}{%
{\protect \APACyear {2019}}%
}]{%
schwartz19}
\APACinsertmetastar {%
schwartz19}%
\begin{APACrefauthors}%
{Schwartz}, S\BPBI J.%
, {Andersson}, L.%
, {Xu}, S.%
, {Mitchell}, D\BPBI L.%
, {Akbari}, H.%
, {Ergun}, R\BPBI E.%
\BDBL {}{Meziane}, K.%
\end{APACrefauthors}%
\unskip\
\newblock
\APACrefYearMonthDay{2019}{{\APACmonth{11}}}{}.
\newblock
{\BBOQ}\APACrefatitle {{Collisionless Electron Dynamics in the Magnetosheath of
  Mars}} {{Collisionless Electron Dynamics in the Magnetosheath of
  Mars}}.{\BBCQ}
\newblock
\APACjournalVolNumPages{\grl}{46}{21}{11,679-11,688}.
\newblock
\begin{APACrefDOI} \doi{10.1029/2019GL085037} \end{APACrefDOI}
\PrintBackRefs{\CurrentBib}

\bibitem [\protect \citeauthoryear {%
{Schwartz}%
, {Daly}%
\BCBL {}\ \BBA {} {Fazakerley}%
}{%
{Schwartz}%
\ \protect \BOthers {.}}{%
{\protect \APACyear {1998}}%
}]{%
schwartz98}
\APACinsertmetastar {%
schwartz98}%
\begin{APACrefauthors}%
{Schwartz}, S\BPBI J.%
, {Daly}, P\BPBI W.%
\BCBL {}\ \BBA {} {Fazakerley}, A\BPBI N.%
\end{APACrefauthors}%
\unskip\
\newblock
\APACrefYearMonthDay{1998}{{\APACmonth{01}}}{}.
\newblock
{\BBOQ}\APACrefatitle {{Multi-Spacecraft Analysis of Plasma Kinetics}}
  {{Multi-Spacecraft Analysis of Plasma Kinetics}}.{\BBCQ}
\newblock
\APACjournalVolNumPages{ISSI Scientific Reports Series}{1}{}{159-184}.
\PrintBackRefs{\CurrentBib}

\bibitem [\protect \citeauthoryear {%
{Schwartz}%
, {Thomsen}%
, {Bame}%
\BCBL {}\ \BBA {} {Stansberry}%
}{%
{Schwartz}%
\ \protect \BOthers {.}}{%
{\protect \APACyear {1988}}%
}]{%
schwartz88}
\APACinsertmetastar {%
schwartz88}%
\begin{APACrefauthors}%
{Schwartz}, S\BPBI J.%
, {Thomsen}, M\BPBI F.%
, {Bame}, S\BPBI J.%
\BCBL {}\ \BBA {} {Stansberry}, J.%
\end{APACrefauthors}%
\unskip\
\newblock
\APACrefYearMonthDay{1988}{{\APACmonth{11}}}{}.
\newblock
{\BBOQ}\APACrefatitle {{Electron heating and the potential jump across fast
  mode shocks}} {{Electron heating and the potential jump across fast mode
  shocks}}.{\BBCQ}
\newblock
\APACjournalVolNumPages{\jgr}{93}{}{12923-12931}.
\newblock
\begin{APACrefDOI} \doi{10.1029/JA093iA11p12923} \end{APACrefDOI}
\PrintBackRefs{\CurrentBib}

\bibitem [\protect \citeauthoryear {%
{Vignes}%
\ \protect \BOthers {.}}{%
{Vignes}%
\ \protect \BOthers {.}}{%
{\protect \APACyear {2000}}%
}]{%
vignes00}
\APACinsertmetastar {%
vignes00}%
\begin{APACrefauthors}%
{Vignes}, D.%
, {Mazelle}, C.%
, {Reme}, H.%
, {Acu{\~n}a}, M\BPBI H.%
, {Connerney}, J\BPBI E\BPBI P.%
, {Lin}, R\BPBI P.%
\BDBL {}{Ness}, N\BPBI F.%
\end{APACrefauthors}%
\unskip\
\newblock
\APACrefYearMonthDay{2000}{{\APACmonth{01}}}{}.
\newblock
{\BBOQ}\APACrefatitle {{The solar wind interaction with Mars: Locations and
  shapes of the bow shock and the magnetic pile-up boundary from the
  observations of the MAG/ER Experiment onboard Mars Global Surveyor}} {{The
  solar wind interaction with Mars: Locations and shapes of the bow shock and
  the magnetic pile-up boundary from the observations of the MAG/ER Experiment
  onboard Mars Global Surveyor}}.{\BBCQ}
\newblock
\APACjournalVolNumPages{\grl}{27}{1}{49-52}.
\newblock
\begin{APACrefDOI} \doi{10.1029/1999GL010703} \end{APACrefDOI}
\PrintBackRefs{\CurrentBib}

\bibitem [\protect \citeauthoryear {%
{Xu}%
\ \protect \BOthers {.}}{%
{Xu}%
\ \protect \BOthers {.}}{%
{\protect \APACyear {2020}}%
}]{%
xu20}
\APACinsertmetastar {%
xu20}%
\begin{APACrefauthors}%
{Xu}, S.%
, {Mitchell}, D\BPBI L.%
, {Schwartz}, S\BPBI J.%
, {Horaites}, K.%
, {Andersson}, L.%
, {Mazelle}, C.%
\BDBL {}{Espley}, J.%
\end{APACrefauthors}%
\unskip\
\newblock
\APACrefYearMonthDay{2020}{}{}.
\newblock
{\BBOQ}\APACrefatitle {{Cross-shock electrostatic potentials at Mars inferred
  from MAVEN measurements}} {{Cross-shock electrostatic potentials at Mars
  inferred from MAVEN measurements}}.{\BBCQ}
\newblock
\APACjournalVolNumPages{\grl, submitted for review}{}{}{}.
\PrintBackRefs{\CurrentBib}

\end{thebibliography}

%Reference citation instructions and examples:
%
% Please use ONLY \cite and \citeA for reference citations.
% \cite for parenthetical references
% ...as shown in recent studies (Simpson et al., 2019)
% \citeA for in-text citations
% ...Simpson et al. (2019) have shown...
%
%
%...as shown by \citeA{jskilby}.
%...as shown by \citeA{lewin76}, \citeA{carson86}, \citeA{bartoldy02}, and \citeA{rinaldi03}.
%...has been shown \cite{jskilbye}.
%...has been shown \cite{lewin76,carson86,bartoldy02,rinaldi03}.
%... \cite <i.e.>[]{lewin76,carson86,bartoldy02,rinaldi03}.
%...has been shown by \cite <e.g.,>[and others]{lewin76}.
%
% apacite uses < > for prenotes and [ ] for postnotes
% DO NOT use other cite commands (e.g., \citet, \citep, \citeyear, \nocite, \citealp, etc.).
%

\end{document}

\title{Observations of Energized Electrons in the Martian Magnetosheath \\ Supplementary Material}

\authors{K.~Horaites\affil{1}, L.~Andersson\affil{1}, S.~J.~Schwartz\affil{1}, S.~Xu\affil{2}, D.~L.~Mitchell\affil{2}, C.~Mazelle\affil{3}, J.~Halekas\affil{4}, J.~Gruesbeck\affil{5}}

 \affiliation{1}{Laboratory for Atmospheric and Space Physics, Boulder, CO, USA}
 \affiliation{2}{Space Sciences Laboratory, Berkeley, CA, USA}
 \affiliation{3}{IRAP CNRS-University of Toulouse-UPS-CNES, Toulouse, France}
 \affiliation{4}{Department of Physics and Astronomy, University of Iowa, Iowa City, IA, USA}
 \affiliation{5}{Goddard Space Flight Center, Greenbelt, Maryland, USA}

\section{Introduction}

This document covers the technical aspects behind the analysis provided in the main paper ``Observations of Energized Electrons in the Martian Magnetosheath''. In section~\ref{collisionless_theory_sec}, the Liouville mapping as it pertains to the present study is reviewed. Section~\ref{theory_delta_phi_sec} develops a prediction for the main quantity of interest, $\Delta \Phi$, as it might be observed just downstream of the Martian bow shock. The discrepancy between this model---which is based on the present understanding of electron energization at Mars---and the observations of $\Delta \Phi$ is the basis of the discussion section of the main paper. Section~\ref{obs_phi_sec} details how the energization $\Phi$ is measured using a weighted fit. The method involves comparing an electron velocity distribution function with a reference distribution sampled in the solar wind. The process used to construct this solar wind reference is described in section~\ref{obs_sw_sec}. Finally, section~\ref{error_sec} explains how the uncertainties used in our weighted fitting procedure are quantified.

\section{Theory}\label{theory_sec}
\subsection{Collisionless Kinetic Theory}\label{collisionless_theory_sec}
In a collisionless plasma, the evolution of the electron distribution function $f(\bvec{v}, \bvec{x}, t)$ obeys {\it Liouville's theorem}. This well-known result, which follows directly from the collisionless Boltzmann equation, dictates that the phase space density advects with the trajectories of particles (here, electrons) in that system. That is, if $[\bvec{x}(t), \bvec{v}(t)]$ is the solution for the motion of a particle with initial conditions $[\bvec{x}(0), \bvec{v}(0)]$, then the phase space density is constant along that trajectory: $f(\bvec{v}(t), \bvec{x}(t), t) = f(\bvec{v}(0), \bvec{x}(0), 0)$.

 %That is, if a representative particle follows some trajectory in position-velocity phase space---say, $\bvec{x}(t)= \tilde{\bvec{x}}(t)$, $\bvec{v(t)} = \tilde{\bvec{v}}(t)$, where $ \tilde{\bvec{x}}(t)$, $ \tilde{\bvec{v}}(t)$ are functions chosen to describe the particle's motion---then the phase space density is found to be constant along that trajectory: $f(\tilde{\bvec{v}}(t), \tilde{\bvec{x}}(t), t)$ = const. 

Liouville's Theorem is particularly useful if one can also find constants of motion that constrain the particle's evolution in phase space. The knowledge of such constants allows one to perform a ``Liouville mapping'' \cite{daly08}, by equating the amplitude of $f$ in two disparate regions of phase space that lie along the same inferred particle trajectory. 
%In this paper, we will consider a simple scenario in which the magnetosheath (in some inertial frame in which Mars is fixed) is steady-state; notably, the bulk velocity $\bvec{v}_{b}(\bvec{x})$, electric field $\bvec{E}(\bvec{x})$, and magnetic field $\bvec{B}(\bvec{x})$, are independent of time.
%If the electromagnetic fields $\bvec{E}(\bvec{x})$ and $\bvec{B}(\bvec{x})$ are slowly varying, we may adopt the common assumption \cite{schwartz98} that throughout their motion the electrons preserve their total energy $\mathcal{E}$ and magnetic moment $M$, quantities which we here explicitly define:
If the electromagnetic fields $\bvec{E}(\bvec{x})$ and $\bvec{B}(\bvec{x})$ are slowly varying, we may adopt the common assumption \cite{schwartz98} that throughout its motion an electron preserves its magnetic moment $M$, which we here explicitly define:

%\begin{equation}\label{E_eq_def}
%\mathcal{E} \equiv \frac{m_e v^2}{2} - \Phi,
%\end{equation}

\begin{equation}\label{M_eq_def}
M \equiv \frac{m_e v_\perp^2}{2 B(\bvec{x})}.
\end{equation}

\noindent In Eq.~(\ref{M_eq_def}) above, $m_e$ refers to the electron mass and $v_\perp$ is the perpendicular component of the electron velocity relative to the magnetic field.

Let us consider the phase space evolution of a particle that is observed in the solar wind at position $\bvec{x}_1$, which then crosses the bow shock before being observed at position $\bvec{x}_2$ in the sheath. \red{We will treat this problem in an inertial frame in which Mars's center of mass is taken to be at rest.} We will assume that in transit the particle with initial velocity $\bvec{v}_1$ gained some kinetic energy while preserving its magnetic moment $M$, and is accelerated to velocity $\bvec{v}_2$ by the time of its arrival at $\bvec{x}_2$. With this picture in mind, let us then relate the initial and final kinetic energies:

\begin{equation}\label{E_eq_def}
%\mathcal{E}\equiv \frac{m_e v_2^2}{2} = \frac{m_e v_1^2}{2} + \Phi,
\frac{m_e v_2^2}{2} = \frac{m_e v_1^2}{2} + \Phi,
\end{equation}

%\noindent where the $\mathcal{E}$ is a label that we will use to refer to the particle energy, and we have introduced the variable ${\Phi}$ to quantify the particle energization provided by the electric field.
\noindent where we have introduced the variable ${\Phi}$ to quantify the particle energization provided by the electric field.
%\noindent where the $\mathcal{E}$ is a label that we will use to refer to the particle energy, and we have introduced the variable ${\Phi}$ to quantify the particle energization provided by the electric field.
 Such energization may arise, e.g., in the presence of an electrostatic potential $\phi(\bvec{x})$, in which case we would identify $\Phi = |q_e| (\phi(\bvec{x}_2) - \phi(\bvec{x}_1))$ where $q_e$ is the electron charge.
% Indeed, in steady state $\bvec{E}(\bvec{x})$ must be the gradient of some scalar potential, as required by Faraday's Law; however, this assumption would ignore the possible influence of electric field fluctuations, as may occur due to kinetic instabilities or turbulence \cite{galeev76}. We do not here account for these mechanisms as they may pertain to electron heating in the shock/sheath system.  Rather, we will conduct our data analysis (section \ref{obs_sec}) with an agnostic view as to the source of electron energization, and use the notation $\Phi$ as shorthand for ``the observed scalar energization of sheath electrons''.
% We will revisit the nature of the electron energization, and of the electric field in the Martian system, in section \ref{discussion_sec}.
The nature of the electron energization and of the electric field in the Martian system is addressed in detail in the discussion section of the main paper. 

Let take as given the particle's initial (in the solar wind) field-parallel and -perpendicular velocity components $v_{\parallel,1}$ and $v_{\perp,1}$, and the initial magnetic field $B_1$. Then, given the final (in the sheath) magnetic field $B_2$, we may calculate the final velocity components $v_{\parallel, 2}$ and $v_{\perp,2}$  from algebraic manipulation of Eqs.~(\ref{M_eq_def})-(\ref{E_eq_def}) :  

\begin{equation}\label{v_perp_eq}
v_{\perp,2} = v_{\perp,1} \sqrt{\frac{B_2}{B_1}},
\end{equation}

\begin{equation}\label{v_par_eq}
%v_{\parallel,2}^2 = v_{\parallel,1}^2 + v_{\perp,1}^2\Big(1 -  \frac{B_2}{B_1}\Big) + \frac{2 |q_e|}{m_e}\phi.
v_{\parallel,2}^2 = v_{\parallel,1}^2 + v_{\perp,1}^2\Big(1 -  \frac{B_2}{B_1}\Big) + \frac{2 \Phi}{m_e}.
\end{equation}

From Eq.~(\ref{v_par_eq}), we may calculate the final pitch angle $\theta_2 = \cos^{-1}(v_{\parallel,2}/v_2)$ of the electron velocity with respect to the magnetic field. Generally, we may expect electrons moving through a spatially varying magnetic field to experience magnetic mirroring; the critical condition associated with mirroring corresponds with a final pitch angle $\theta_2 = 90^\circ$, or alternatively, $v_{\parallel,2} =0$. From setting $v_{\parallel,2} = 0$ in Eq.~(\ref{v_par_eq}) we arrive at a formula for the pitch angle $\theta_{1,c}$ defining the ``loss cone'':
%\footnote{The shape of the ``loss cone'' (in velocity space) in a system with spatially varying potential is technically not conical, due to the presence of the rightmost (energy-dependent) term in the parentheses of Eq.~(\ref{loss_cone_eq}).}:

\begin{equation}\label{loss_cone_eq}
%\theta_{1,c} = \sin^{-1} \sqrt{\frac{B_1}{B_2} \Big( 1 + \frac{2 |q_e| \phi}{m_e v_1^2}\Big)}.
\theta_{1,c} = \sin^{-1} \sqrt{\frac{B_1}{B_2} \Big( 1 + \frac{2 \Phi}{m_e v_1^2}\Big)}.
\end{equation}

\noindent If the initial pitch angle $\theta_1$ satisfies $\theta_1>\theta_{1,c}$, then the particle will be reflected before it reaches the sheath location $\bvec{x}_2$. Some care must therefore be taken in the process of Liouville mapping, to identify which portions of the distribution function represent reflected particles, as in e.g. \cite{knight73}.

We now note that a particle originating in the solar wind that is exactly aligned (or anti-aligned) with the magnetic field will never deviate in pitch angle, regardless of the particular profile of the magnetic field $B(\bvec{x})$. We see this follows directly from Eq.~(\ref{v_perp_eq}), as we must have $v_{\perp,2}=0$ if  $v_{\perp,1}=0$. We therefore assert that the parallel and anti-parallel cuts of the distribution function $f$---denoted here (as a function of energy) as $f_\parallel(v^2)$ and $f_\downarrow(v^2)$ respectively---are particularly suited for analysis via Liouville mapping. That is, if a parallel cut $f_{1,\parallel}$ is observed in the solar wind, we expect (from Eq.~\ref{E_eq_def}) the parallel sheath cut $f_{2,\parallel}$ to appear identical up to a constant energy shift $\Phi_\parallel$:

\begin{equation}\label{liouville_par_eq}
f_{2,\parallel}\Big(v^2 + \frac{2 \Phi_\parallel}{m_e}\Big) = f_{1,\parallel}(v^2).
\end{equation}

\noindent We here assume that all electrons comprising $f_{2,\parallel}$ were energized by the same amount, as may be justified a posteriori from inspection of the data. We may express the Liouville mapping for antiparallel cut ($f_{\downarrow}$) by an analogous formula, with $\Phi_\downarrow$ denoting the energy shift. We note that Eq.~(\ref{liouville_par_eq}) tells us nothing about the distribution $f_{2,\parallel}(v^2)$ at kinetic energies less than $\Phi_\parallel$. Indeed, this void in phase space cannot be populated by electrons originating from our example location $\bvec{x}_1$. In practice, this region of phase space typically exhibits a ``flat top'' ($f=$const.) shape in Mars's magnetosheath \cite{crider00}; electron distributions in Earth's magnetosheath exhibit a similar feature \cite{feldman83a}.

\subsection{Field-Parallel Anisotropy $\Delta \Phi$}\label{theory_delta_phi_sec}
Let us consider a steady-state model of the Martian system, closely related to the framework adopted in \cite{schwartz19}. That is, we assume the distribution function in the solar wind $f_{sw}(\bvec{v})$ is prescribed and is independent of position. We assume that electrons populating the Martian magnetosheath have entered by traveling along a magnetic field line that connects back out to the solar wind. We take the bulk flow of the solar wind plasma to be constant outside the bow shock $\bvec{v}_{sw}=-\hat \bvec{x}_{mse}v_{sw}$, with some deceleration profile within the sheath. Note that throughout this paper the subscript ``mse'' is used to denote coordinates in the Mars Solar Electric (MSE) frame.  \red{We will specify this frame in Cartesian coordinates in terms of the conventional unit vectors $\hat \bvec{x}_{mse}$, $\hat \bvec{y}_{mse}$, $\hat \bvec{z}_{mse}$, using Mars's center as the frame's origin: $\hat \bvec{x}_{mse} \equiv -(\bvec{v}_{sw} / v_{sw})$ points opposite the incoming solar wind velocity $\bvec{v}_{sw}$, $\hat \bvec{z}_{mse} \equiv -\bvec{v}_{sw}\times\bvec{B} / |\bvec{v}_{sw}\times\bvec{B}|$ is parallel to the upstream convective electric field, and $\hat \bvec{y}_{mse} \equiv \hat \bvec{z}_{mse} \times \hat \bvec{x}_{mse}$ completes the orthonormal set of unit vectors. Note that under this definition, $\bvec{B}\cdot\hat\bvec{y}_{mse}$$>$0.}

The magnetic field, which is frozen into the decelerating plasma, should then ``drape'' \cite{crider04} around Mars as the solar wind encounters the planet; this effect can also be seen in MHD simulations \cite{luhmann15}. Outside the bow shock in the solar wind the magnetic field lines are taken to be straight, with some angle of inclination $\theta_c$ with respect to the vector $\bvec{v}_{sw}$:

\begin{equation}\label{theta_c_eq}
\theta_c \equiv \cos^{-1}(\bvec{B} \cdot \bvec{v}_{sw} / B v_{sw}).
\end{equation}

\noindent Under steady conditions we may expect the ``cone angle'' $\theta_c$ to be given by the Parker spiral angle; i.e. $\theta_{c}\approx 60^\circ$ at Mars.
In our primary reference \cite{schwartz19}, an inclination angle $\theta_c$=90$^\circ$ was adopted; we relax that assumption here to more realistically model the solar wind.
%Indeed, in the case $\theta_c$=90$^\circ$ we predict an isotropic energization ($\Delta \Phi$=0) due to the symmetry of the system.
 With this framework in mind, we note that a field line that crosses into the sheath should pass through the bow shock at two ends, and the line segment connecting these points will have the same inclination angle $\theta_c$.

The average shape of the Martian bow shock was measured in \cite{vignes00}. Following \cite{schwartz19}, we will approximate the curved shock surface with the parabolic form:

\begin{equation}\label{bow_shock_eq}
a r^2 = (R_B - x). 
\end{equation}

\noindent Equation~(\ref{bow_shock_eq}) assumes an inertial frame where Mars's center lies at the origin; the variable $x$ represents the distance along the Mars-Sun line and the cylindrical coordinate $r$ represents the distance from the $x$-axis. We also introduce the constants $a = 0.21 R_{m}^{-1}$ and $R_B = 1.645 R_{m}$ to define the bow shock's parabolic shape, where $R_{m}=3389.5$km is the nominal Martian radius.

In typical descriptions of the Martian magnetosheath, the last element of the physical ``framework'' relevant to our observations is the prescription of a cross-shock potential $\Phi_s$ that is wholly responsible for the energization of the electrons.  The influence of this shock potential is confined to the local region of the shock itself. In the typical picture the shock potential arises from a \red{frame-invariant} ambipolar electric field established over a thin region aligned with the shock front, whose purpose is to confine electrons and maintain quasineutrality. The spatial variation of this cross-shock potential may be determined from the observed energization of the electrons. \red{Following early observations that the shock potential is proportional to the incident ram energy \cite{schwartz88}, we here assume that for a given ram energy this potential is dependent on the shock angle $\theta_{vn}$:}
%a recent empirical study of the Martian bow shock \cite{xu20} demonstrates that for a given ram energy this potential is dependent on the shock angle $\theta_{vn}$:}

\begin{equation}\label{theta_vn_eq}
\theta_{vn}\equiv \cos^{-1} ( - \hat \bvec{v}_{sw}\cdot \hat \bvec{n}).
\end{equation}

\noindent In Eq.~(\ref{theta_vn_eq}), $ \hat \bvec{v}_{sw}= \bvec{v}_{sw}/ v_{sw}$ and $\hat \bvec{n}$ is the unit vector directed normal to the shock surface ($\hat \bvec{n}$ points away from the sheath and towards solar wind). Note that $0^\circ$ $<$ $\theta_{vn}$ $<$ $90^\circ$ for the parabolic form of the bow shock assumed in Eq.~\ref{bow_shock_eq}.  We may now assume $\Phi_s = \Phi_s(\theta_{vn})$, we may adopt the particular form of this function:

\begin{equation}\label{phi_s_eq}
\Phi_s(\theta_{vn}) = \Phi_{s,0} \cos^2(\theta_{vn}),
\end{equation}

%\noindent where $\Phi_{s,0}\sim$100 eV is the peak potential, located on the bow shock at the subsolar point (where $\theta_{vn}=$0). The shock angle itself is defined as the angle between the shock normal $\hat \bvec{n}$ and the $\bvec{x}_{mse}$-axis:  $\theta_{vn} \equiv \cos^{-1} ( \hat \bvec{n}\cdot  \hat \bvec{x}_{mse})$. Assuming our model of the bow shock shape, Eq.~(\ref{bow_shock_eq}), we may express the shock angle $\theta_{vn}$ in terms of the solar zenith angle $\theta_{sz}$: \comment{FOR WHAT $\theta_{sz}$ IS THIS VALID?}

\noindent where $\Phi_{s,0}\sim$100 eV is the peak potential, located on the bow shock at the subsolar point (where $\theta_{vn}=$0). The variation of $\Phi_s(\theta_{vn})$ is readily interpreted as being proportional to the solar wind ram energy density, $\rho v_n^2 / 2$, where $\rho$$\sim$const. is the solar wind mass density immediately outside the bow shock and $v_n = v_{sw}\cos \theta_{vn}$. \red{The empirical demonstration of this proportionality using MAVEN data is the topic of ongoing work.}

We arrive at a key point: the field-parallel and anti-parallel electrons observed at a given location in the sheath will have entered by crossing the bow shock at opposite ends of the same field line (with different $\theta_{vn}$), and therefore in general {\it they will have been energized by different shock potentials}. That is, we expect the observed anisotropy of the energization as seen in the sheath to be non-zero: 

\begin{equation}\label{delta_phi_def_eq}
\Delta \Phi = \Phi_\parallel - \Phi_\downarrow \neq 0.
\end{equation}

\noindent Let us quantify the anisotropy $\Delta \Phi$ as predicted from our model of the shock potential (Eqs.~\ref{phi_s_eq}, \ref{theta_sh_eq}) for a given solar wind cone angle $\theta_c$. Since the field lines are frozen into the plasma flow, the locations where the electrons crossed the shock may be determined geometrically from the instantaneous fields as described above. We will only consider the plane $z_{mse}=0$, in which the out-of-plane component of the magnetic field is expected to be zero. 
In this plane, let us introduce the polar angle $\theta_{xy}$:           %$\theta_{xy} = \tan^{-1} (y_{mse}/ x_{mse})$:

\begin{equation}\label{theta_xy_eq}
%\theta_{xy} = \text{{\bf atan2}} (y_{mse}, x_{mse}).
\theta_{xy} = \mathrm{{\bf atan2}} (y_{mse}, x_{mse}).
\end{equation}

\pagebreak

\noindent Note -180$^\circ$$<$$\theta_{xy}$$<$180$^\circ$. Near the shock front in the $z_{mse}=0$ plane, $\Delta \Phi$ is a function depending on the variable $\theta_{xy}$ and the parameters $\theta_c$, $\Phi_{s,0}$:

\begin{equation}\label{delta_phi_dep_eq}
\Delta \Phi = \Delta\Phi(\theta_{xy}; \theta_c, \Phi_{s,0}) \mathrm{, \hspace{6pt}if \hspace{3pt}} z_{mse}=0.
\end{equation}

%Assuming our model of the bow shock shape, Eq.~(\ref{bow_shock_eq}), we may express the shock angle $\theta_{vn}$ in terms of the solar zenith angle $\theta_{sz}$: \comment{FOR WHAT $\theta_{sz}$ IS THIS VALID?}
%Let us consider z=0 plane...
Assuming our model of the bow shock shape, Eq.~(\ref{bow_shock_eq}), we may express the shock angle $\theta_{vn}$ in terms of the azimuthal angle $\theta_{xy}$:

\begin{equation}\label{theta_sh_eq}
%\theta_{vn}(\theta_{xy}) = \cot^{-1}\Big\{ -\cot (\theta_{xy})  + \sqrt{\cot^2(\theta_{xy}) + 4 a R_B}           \Big\}.
\theta_{vn}(\theta_{xy}) = \tan^{-1}\Big\{ -\cot |\theta_{xy}|  + \sqrt{\cot^2(\theta_{xy}) + 4 a R_B} \Big\}.
\end{equation}

\noindent We may combine Eqs.~(\ref{phi_s_eq}),(\ref{theta_sh_eq}) to obtain a formula for the shock potential as a function of the azimuthal angle, $\Phi_s(\theta_{xy})$:

\begin{equation}\label{phi_s_eq2}
\Phi_s(\theta_{xy}) = \Phi_{s,0} \Big\{1 + \Big[ -\cot|\theta_{xy}| + \sqrt{\cot^2(\theta_{xy}) + 4 a R_B} \Big]^2 \Big\}^{-1}
\end{equation}

\noindent In our framework, computing the difference between two values of $\Phi_s(\theta_{xy})$ at two shock crossing points along the same field line can then be used to predict $\Delta \Phi$ just downstream of those crossing points. 

%We may combine Eqs.~(\ref{phi_s_eq}),(\ref{theta_sh_eq}) to arrive at a formula for the shock potential as a function of solar zenith angle, $\Phi_s(\theta_{sz})$. 
%We arrive at a key point: the field-parallel and anti-parallel electrons observed at a given location in the sheath will have entered by crossing the bow shock at opposite ends of the same field line (with different $\theta_{sz}$), and therefore in general {\it they will have been energized by different shock potentials}. That is, we expect the observed anisotropy of the energization as seen in the sheath to be non-zero: $\Delta \Phi \neq 0$.  To quantify this claim, let us calculate the anisotropy $\Delta \Phi$ as predicted from our model of the shock potential (Eqs.~\ref{phi_s_eq}, \ref{theta_sh_eq}) for a given solar wind cone angle $\theta_c$, For now we ignore the advection of the magnetic field lines and assume that the locations where the electrons crossed the shock may be determined geometrically from the instanteous fields as described above. We will also only consider the plane $z_{mse}=0$, in which the out-of-plane component of the magnetic field is expected to be zero. It then follows that the potential anisotropy $\Delta \Phi$, near the shock front in the $z_{mse}=0$ plane, should only depend on the polar angle $\theta_{xy} = \tan^{-1} (y_{mse}/ x_{mse})$, cone angle $\theta_c$, and peak potential $\Phi_{s,0}$ :

%\begin{equation}
%\Delta \Phi = \Delta\Phi(\theta_{xy}; \theta_c, \Phi_{s,0}) \text{, \hspace{6pt}if \hspace{3pt}} z_{mse}=0.
%\end{equation}

\begin{figure}
\includegraphics[width=1\linewidth]{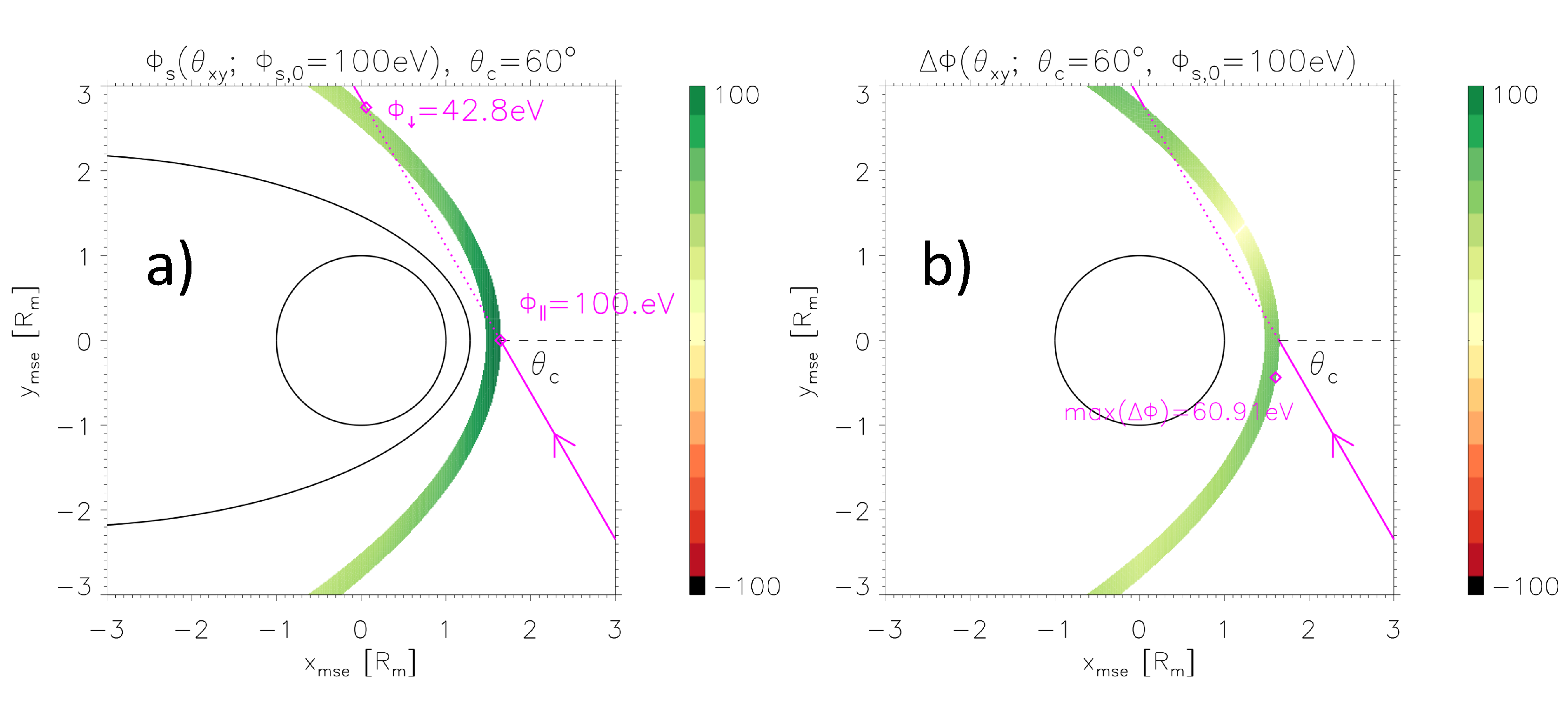}
\caption{\label{phi_anisotropy_model_fig} 
\red{ {\bf a.} The model cross-shock potential $\Phi_s(\theta_{xy})$ assumed in this paper (in eV), assuming $\Phi_{s,0}$=100 eV. The formula is given by Eq.~\ref{phi_s_eq2}, following the observational trend measured in \citeA{xu20}. A magnetic field line with cone angle $\theta_c$=60$^\circ$ is shown as a pink line (note the anti-sunward orientation). This field line crosses the model bow shock (Eq.~\ref{bow_shock_eq}) at two points marked by diamonds: the nose of the shock ($\theta_{xy}$=0$^\circ$) and another point on the flank of the shock. The quantities $\Phi_\parallel$=100eV and $\Phi_\downarrow$=42.8eV are shown at the two crossing points, from Eq.~\ref{phi_s_eq2}. As electrons propagate quickly enough that we may approximate the field lines are essentially fixed, a picture in which electron energization solely occurs at the shock leads to the conclusion that $\Phi_\parallel$ and $\Phi_\downarrow$ are mirrored at the two shock crossing points of a given field line. Therefore our model predicts ${\Delta\Phi=(\Phi_\parallel-\Phi_\downarrow})$=57.2eV at both of the marked points.}
{\bf b.} The predicted $\Delta \Phi(\theta_{xy})$ for the case $\theta_c$=60$^\circ$, $\Phi_{s,0}$=100eV (see Eqs.~\ref{delta_phi_dep_eq}, \ref{phi_s_eq2}). Our model for $\Delta \Phi(\theta_{xy})$ predicts this quantity just downstream of the Martian bow shock, so the results are plotted as a function of bow shock position in the figure above. To predict $\Delta \Phi$, we compute the difference between two values of $\Phi_s$ at two shock crossing points along the same field line with cone angle $\theta_c$ (pictured); this yields $\Delta \Phi$ just downstream of those crossing points. We iteratively propagate the field line in along the $x_{mse}$-axis and repeat this procedure to map $\Delta \Phi$ along the entire shock. For this case, where the magnetic field is directed anti-sunward and $\Phi_s(\theta_{xy})$ falls off monotonically with $|\theta_{xy}|$, we find ${\Delta\Phi>0}$. We predict a maximum energization differential $\max(\Delta\Phi)$$\approx$60eV near the subsolar point (${\theta_{xy}=0}$).}
%is predicted directly behind the shock in the sheath, which arises because each magnetic field line crosses the shock at disparate locations. We assume a solar wind cone angle $\theta_c =$60$^\circ$, the typical value seen at Mars.}
\end{figure}

In Figure~\ref{phi_anisotropy_model_fig} we plot the potential anisotropy $\Delta\Phi(\theta_{xy}; \theta_c, \Phi_{s,0})$ that follows from Eqs.~(\ref{phi_s_eq}),(\ref{delta_phi_dep_eq}),(\ref{theta_sh_eq}) assuming $\theta_c$=60$^\circ$ and $\Phi_{s,0}$=100 eV. We generate this plot by propagating a field line with fixed cone angle $\theta_c$ along the $x_{mse}$-axis, assuming the $B$-field is oriented anti-sunwards; the two points at which this line intersects the model bow shock (from Eq.~\ref{bow_shock_eq}) then respectively yield $\Phi_\parallel$ and $\Phi_\downarrow$. The difference $\Delta \Phi = \Phi_\parallel - \Phi_\downarrow$ is then calculated, which gives the potential anisotropy at the points where the field line intersects the bow shock. As an aside, we note that this same anisotropy $\Delta \Phi$ would be expected everywhere along the same field line, not just at its intersection with the shock, but the field line's curved path within the sheath is not known a priori because we do not specify the spatial profile of the bulk flow in this region.  We now note some important features of the predicted anisotropy displayed in Fig.~\ref{phi_anisotropy_model_fig}: 1. the potential difference $\Delta \Phi$ is strictly $\geq$0 (because $\bvec{B}$ points antisunward in the solar wind), 2. $\Delta \Phi$ exhibits a maximum near the subsolar point, with a peak value of $\sim$0.61$\Phi_{s,0}=$61eV, and 3. $\Delta \Phi$ vanishes at the point where the solar wind magnetic field lies tangent to the shock surface ($\bvec{B} \cdot \hat \bvec{n} $=0).

\section{Methodology}
\subsection{Calculating $\Phi$}\label{obs_phi_sec}

%energy truncation and identification of pitch angle branches

We here explain the process for calculating the energization ($\Phi$) of a single electron velocity distribution cut measured by MAVEN. We will apply this process for every distribution in our data set, and obtain values for the parallel and anti-parallel energizations, $\Phi_\parallel$ and $\Phi_\downarrow$, respectively. The difference between these quantities, $\Delta \Phi$, will be the parameter of interest in our global characterization of field-parallel anisotropy in the sheath (see main paper).

The first step of the process is to obtain parallel and anti-parallel cuts of the electron distribution. Given a discretely sampled spectrum (32 energies, 16 pitch angles) measured by the SWEA instrument, we first identify angular branches for which the pitch angle $\theta$ is most near to $0^\circ$ and $180^\circ$, which defines the parallel and anti-parallel cuts respectively. We only consider cases for which both branches are sufficiently aligned to the magnetic field, i.e. within $30^\circ$. \comment{(***REFER to Eq.~\ref{loss_cone_eq} here?)} At lower energies the sampled distribution may be contaminated by photoelectrons or warped by the effects of spacecraft potential,\comment{(CITATION***)} and higher energies frequently exhibit very low count rates---therefore, we explicitly only retain bins of the spectrum that register $\geq$5 counts, and we furthermore only consider energies in the range 30-500 eV. Within this energy range, we then identify the peak of the corresponding cut of the differential energy flux. The differential energy flux $D(\bvec{v})$ is simply related to the distribution $f(\bvec{v})$ by the proportionality $D(\bvec{v}) \propto v^4 f(\bvec{v})$; in the sheath the energy $\Phi_{est}$ at the peak of $D(v)$ may be used as an estimate of the energization $\Phi$ (for an energy cut of the distribution), as the peak typically occurs at an energy where the low-energy flat-top in $f_\parallel(v)$ begins to fall off. \comment{(CITATION***)} In addition to the low-energy cutoff at 30 eV, we only consider bins of the distribution at energies $\geq$$\Phi_{est}$; this excludes the flat-top portion from our fitting procedure.

%interpolation

Equation (\ref{liouville_par_eq}), which is at the heart of our Liouville mapping procedure, requires us to compare two distributions {\it at the same phase space density (PSD)} in order to derive $\Phi$. The second step of our procedure is then to interpolate our discretely sampled data, to infer the energy at which the spectra in the solar wind and sheath would be expected to have the same PSD. We therefore apply a linear interpolation between the solar wind and magnetosheath spectra, which we will refer to with the subscripts $w$ and $m$ respectively. Let us consider a data point in our cut of the solar wind distribution, with an observed PSD $f_w$ measured at energy $\mathcal{E}_w$.  We then identify two bins in the magnetosheath cut, with PSD $f_{m2}$ and $f_{m1}$ respectively, that are so chosen as to minimize the differences $|f_{m2}-f_w|$, $|f_{m1}-f_w|$ while satisfying the conditions $f_{m1} > f_w$ and $f_{m2} \leq f_w$. We will denote the energies corresponding to these two ``nearest neighbors'' with PSD $f_{m1}$, $f_{m2}$ (which bound $f_w$) as $E_{m1}$, $E_{m2}$ respectively. Then, we may calculate the interpolated energy $E_{m,in}$ at which the magnetosheath spectrum would have PSD of $f_w$:

\begin{equation}\label{E_interp_eq}
%E_{m,in} = E_{m2} + \Big( \frac{E_{m2} - E_{m1}}{f_{m2} - f_{m1}} \Big)(f_w - f_{m2}).
\mathcal{E}_{m,in} = \mathcal{E}_{m2} + (\mathcal{E}_{m1} - \mathcal{E}_{m2})\Big( \frac{f_w - f_{m2}}{f_{m1} - f_{m2}} \Big).
\end{equation}

\noindent We apply this interpolation process (Eq.~\ref{E_interp_eq}) for every datum in the solar wind cut.
% provided suitable nearest neighbors in the magnetosheath cut can be found.
This yields an interpolated magnetosheath cut, where each data point is matched in according to its PSD to a point in the solar wind cut, but these points will be at different energies.
% Measuring the difference in energy at constant amplitude will allow us to estimate $\Phi$ according to  Eq.~(\ref{liouville_par_eq}), as explained below.

%least squares method

The final step in our procedure is then to fit for the energization $\Phi$, by measuring the difference in energy separating the two spectra (Eq.~\ref{liouville_par_eq}), at constant amplitude. Let us denote the energies of these amplitude-matched points, of the solar wind and interpolated magnetosheath spectra, as $\mathcal{E}_{w,j}$ and $\mathcal{E}_{m,in,j}$ respectively. Here we use the letter $j$, where $1 \leq j \leq N$, to index the $N$ data points in each spectrum. Let us denote the energy difference between each pair of points as $\Delta \mathcal{E}_j \equiv (\mathcal{E}_{w,j} - \mathcal{E}_{m,in,j})$. We wish to find the constant shift in energy $\Phi$ required to minimize the difference between the energized solar wind and observed magnetosheath spectra, in the least-squares sense. To this end we define the $\chi^2$ statistic:

\begin{equation}\label{chi_squared_eq}
%\chi^2 \equiv \sum_{j=1}^{N} \frac{(\Delta \mathcal{E}_j + |q_e|\phi)^2}{\sigma^2_{\Delta \mathcal{E},j}}
\chi^2(\Phi) \equiv \sum_{j=1}^{N} \frac{(\Delta \mathcal{E}_j + \Phi)^2}{\sigma^2_{\Delta \mathcal{E},j}}
\end{equation}

\noindent The formula we use to estimate the error $\sigma_{\Delta \mathcal{E},j}$ (standard error of the quantity $\Delta \mathcal{E}_j$) is given in section~\ref{error_sec}. Setting $d\chi^2/d\Phi = 0$ yields a formula for the energization $\Phi$ that minimizes $\chi^2$:

\begin{equation}\label{phi_eq}
\Phi = \frac{ - \sum_{j=1}^{N} ( \Delta \mathcal{E}_j / \sigma^2_{\Delta \mathcal{E},j})}{  \sum_{j=1}^{N} (1 / \sigma^{2}_{\Delta \mathcal{E},j} )}
\end{equation}

\red{\noindent Goodness of fit can be tested using the reduced chi-squared statistic $\chi^2_R$$\equiv$$\chi^2/(N-1)$, see Eq.~\ref{chi_squared_eq}. We require $\chi^2_R$$<$8 for a fit to be included in the study.}

We emphasize that Eq.~\ref{phi_eq} is used to separately calculate the energization of the parallel and anti-parallel cuts, yielding the quantities $\Phi_\parallel$ and $\Phi_\downarrow$, respectively. The standard deviation of this derived value of $\Phi$, which we denote as $\sigma_\Phi$, can be calculated from Eq.~(\ref{phi_eq}) by the technique of propagation of errors:

\begin{equation}\label{sigma_phi_eq}
\sigma_\Phi = \Big( \sum_{j=1}^N \frac{1}{\sigma^2_{\Delta \mathcal{E}, j}}\Big)^{-1/2}.
\end{equation}

\noindent We note that Eq.~(\ref{phi_eq}) can be recognized from statistics as the ``weighted mean'' of the quantity $\Delta \mathcal{E}_j$, while Eq.~(\ref{sigma_phi_eq}) is termed the ``uncertainty of the weighted mean'' \cite{barlow89}.

%%%Example cuts and fit to phi here?

\subsection{Solar Wind Reference Distribution}\label{obs_sw_sec}
%%% WOULD THIS BE BETTER SUITED FOR AN APPENDIX?

% bow shock model

%The method described in section \ref{obs_phi_sec} yields the energization $\Phi$ required to map a parallel (or anti-parallel) cut of the sheath distribution to a corresponding cut of a solar wind reference distribution. This solar wind reference must of course be selected at a time when the MAVEN spacecraft is well beyond Mars's bow shock. For the purpose of selecting suitable solar wind reference distributions, we compare the position of the MAVEN spacecraft with a model of the average bow shock measured in \cite{vignes00}. Following \cite{schwartz19}, we will adopt a parabolic form for the bow shock surface:
The method described in section~\ref{obs_phi_sec} yields the energization $\Phi$ required to map a parallel (or anti-parallel) cut of the sheath distribution to a corresponding cut of a solar wind reference distribution. This solar wind reference must of course be selected at a time when the MAVEN spacecraft is well beyond Mars's bow shock. For the purpose of selecting suitable solar wind reference distributions, we we will consider a 10-minute interval within each orbit when the spacecraft is farthest (in the radial sense) from the average bow shock position (Eq.~\ref{bow_shock_eq}).

% measured in \cite{vignes00}. Following \cite{schwartz19}, we will adopt a parabolic form for the bow shock surface:

%\begin{equation}\label{bow_shock_eq}
%a r^2 = (R_B - x). 
%\end{equation}

%\noindent Equation~(\ref{bow_shock_eq}) assumes an inertial frame where Mars's center lies at the origin; the variable $x$ represents the distance along the Mars-Sun line and the cylindrical coordinate $r$ represents the distance from the $x$-axis. We also introduce the constants $a = 0.21 R_{m}^{-1}$ and $R_B = 1.645 R_{m}$ to define the bow shock's parabolic shape, where $R_{m}=3389.5$km is the nominal Martian radius.

% averaging over minutes of data to get f_sw

For each $\sim$4.5 hour elliptical orbit completed by MAVEN, we conduct the following test to select the data used as our solar wind reference distribution: for each spectrum measured by the SWEA instrument, we find the spacecraft's position $\bvec{x}_{sc}$ in the MSO reference frame and then draw a line from the spacecraft to the origin (Mars's center). Then we find the position $\bvec{x}_{bs}$ at which this line intersects the nominal bow shock surface given Eq.~(\ref{bow_shock_eq}). We then calculate the ratio $|\bvec{x}_{sc}| / |\bvec{x}_{bs}|$; let us denote the time at which this ratio is maximized for that orbit as $t_{w}$. We then compute an average of the survey spectra (described in the main paper), over a 10-minute interval centered on $t_w$; the resulting averaged spectrum constitutes a single solar wind reference, from which for example parallel and anti-parallel cuts can be isolated, to be used in the fitting procedure.

\pagebreak

In the over 4-year interval spanned by our data, MAVEN completed thousands of orbits that extended into the solar wind. For each spectrum for which we calculate $\Phi$ (section \ref{obs_phi_sec}), measured say at time $t_s$, we use the solar wind reference at time $t_w$ that minimizes the difference $|t_s - t_w|$.

\subsection{Estimating Error $\sigma_{\Delta \mathcal{E}, j}$}\label{error_sec}

The fitting procedure used to find the energization $\Phi$ (Eq.~\ref{phi_eq}) and its associated uncertainty $\sigma_\Phi$ (Eq.~\ref{sigma_phi_eq}), described in section~\ref{obs_phi_sec}, requires an estimate of the quantity $\sigma_{\Delta \mathcal{E}, j}$. This quantity is the uncertainty of the energy difference $\Delta \mathcal{E}_j$, between a pair of points (index $j$) from two different spectra with matching phase space density (PSD). Let us here dispense with the subscript $j$, and find the error $\sigma_{\Delta \mathcal{E}}$ of the quantity

\begin{equation}\label{delta_E_app_eq}
\Delta \mathcal{E} = (\mathcal{E}_w - \mathcal{E}_{m,in}).
\end{equation}

\noindent That is, $\sigma_{\Delta \mathcal{E}}$ is the error of the energy difference between an energy bin $\mathcal{E}_w$ of the solar wind spectrum and the interpolated energy bin of the magnetosheath spectrum $E_{m,in}$ (Eq.~\ref{E_interp_eq}). Besides dropping the $j$ subscript, we will otherwise adopt the same notation used in section~\ref{obs_phi_sec}, and write out the formula for $\sigma_{\Delta \mathcal{E}}$ explicitly.

The error we are interested in, $\sigma_{\Delta \mathcal{E}}$, depends on the respective errors $\sigma_{f,w}$, $\sigma_{\mathcal{E},w}$ of the quantities $f_w$, $\mathcal{E}_w$ obtained from the solar wind spectrum (Eq.~\ref{E_interp_eq}). Similarly, we will denote the errors of the quantities $f_{m1}$, $f_{m2}$, $\mathcal{E}_{m1}$, $\mathcal{E}_{m2}$ respectively as $\sigma_{f,m1}$, $\sigma_{f,m2}$, $\sigma_{\mathcal{E},m1}$, $\sigma_{\mathcal{E},m2}$. With this notation, we may write the formula for $\sigma_{\Delta \mathcal{E}}$ which follows from applying propagation of errors to Eq.~\ref{delta_E_app_eq}:

%\begin{split}
\begin{equation}\label{sigma_delta_E_eq}
%\sigma_{\Delta \mathcal{E}}& = \Big\{\sigma^2_{\mathcal{E},w} + \sigma^2_{\mathcal{E},m2} +... \\
%          & \|B\| \sqrt{G(E_{m1}, E_{m2}, \sigma_{E_{m1}}, \sigma_{E_{m2}}) + 
%                                                                      G(f_{m1}, f_{m2}, \sigma_{f_{m1}}, \sigma_{f_{m2}}) + 
%                                                                      G(f_{w}, f_{m2}, \sigma_{f_{w}}, \sigma_{f_{m2}})  } \Big\}^{1/2}
\sigma_{\Delta \mathcal{E}} = \left\{\sigma^2_{\mathcal{E},w} + \sigma^2_{\mathcal{E},m2} + |G| \sqrt{ \frac{\sigma^2_{\mathcal{E},m1} + \sigma^2_{\mathcal{E},m2}}{(\mathcal{E}_{m1}-\mathcal{E}_{m2})^2} + 
%\sigma_{\Delta \mathcal{E}}& = \left\{\sigma^2_{\mathcal{E},w} + \sigma^2_{\mathcal{E},m2} + |B| \sqrt{ \frac{\sigma^2_{\mathcal{E},m1} + \sigma^2_{\mathcal{E},m2}}{(E_{m1}-E_{m2})^2} + 
                                                                         \frac{\sigma^2_{f,m1} + \sigma^2_{f,m2}}{(f_{m1}-f_{m2})^2} +
                                                                         \frac{\sigma^2_{f,w} + \sigma^2_{f,m2}}{(f_{w}-f_{m2})^2}     } \right\}^{1/2},
%          & \|B\| \sqrt{G(E_{m1}, E_{m2}, \sigma_{E_{m1}}, \sigma_{E_{m2}}) + 
%                                                                      G(f_{m1}, f_{m2}, \sigma_{f_{m1}}, \sigma_{f_{m2}}) + 
%                                                                      G(f_{w}, f_{m2}, \sigma_{f_{w}}, \sigma_{f_{m2}})  } \Big\}^{1/2}
\end{equation}
%\end{split}

\noindent where in Eq.~(\ref{sigma_delta_E_eq}) we introduced the quantity $G$ for convenience:

\begin{equation}\label{G_eq}
G\equiv (\mathcal{E}_1 - \mathcal{E}_2)\frac{f_w - f_2}{f_1 - f_2}.
\end{equation}

In order to evaluate $\sigma_{\Delta \mathcal{E}}$, we require an estimate of the of the various errors appearing on the right-hand side of Eq.~(\ref{sigma_delta_E_eq}). The energy errors are straightforward to evaluate, as the SWEA detector specified energies with a full width at half maximum (FWHM) resolution of 17\%. We therefore introduce the function $\sigma_\mathcal{E}(\mathcal{E})$, which denotes the standard error of the energy measured at nominal energy $\mathcal{E}$ by SWEA:

\begin{equation}
\sigma_\mathcal{E}(\mathcal{E}) = \frac{0.17}{\sqrt{8\ln 2}}\mathcal{E}.
\end{equation}

\noindent We then define the energy errors appearing in Eq.~\ref{sigma_delta_E_eq} as: ${\sigma_{\mathcal{E},m1}\equiv \sigma_\mathcal{E}(\mathcal{E}_{m1})}$, ${\sigma_{\mathcal{E},m2}\equiv \sigma_\mathcal{E}(\mathcal{E}_{m2})}$, ${\sigma_{\mathcal{E},w}\equiv \sigma_\mathcal{E}(E_{w})}$.

An estimate of the uncertainty in phase space density (PSD) is also required to evaluate Eq.~(\ref{sigma_delta_E_eq}). For this purpose we will assume that the SWEA detector obeys Poisson (counting) statistics. That is, if a given bin of a SWEA spectrum registers a PSD of $f$ from $C$ individual particle counts, then the associated error is given by the function $\sigma_f(f,C)$:

\begin{equation}\label{poisson_err_eq}
\sigma_f(f,C) = \frac{f}{\sqrt{C}}.
\end{equation}

\noindent The quantities $\sigma_{f,m1}$, $\sigma_{f,m2}$, and $\sigma_{f,w}$ appearing in Eq.~(\ref{sigma_delta_E_eq}) are calculated from the SWEA data ($f_{m1}$, $f_{m2}$, $f_w$ and associated counts) according to Eq.~(\ref{poisson_err_eq}).\comment{may need to evaluate $f_w$ just as SD of averaged data}

\bibliography{maven}

%Reference citation instructions and examples:
%
% Please use ONLY \cite and \citeA for reference citations.
% \cite for parenthetical references
% ...as shown in recent studies (Simpson et al., 2019)
% \citeA for in-text citations
% ...Simpson et al. (2019) have shown...
%
%
%...as shown by \citeA{jskilby}.
%...as shown by \citeA{lewin76}, \citeA{carson86}, \citeA{bartoldy02}, and \citeA{rinaldi03}.
%...has been shown \cite{jskilbye}.
%...has been shown \cite{lewin76,carson86,bartoldy02,rinaldi03}.
%... \cite <i.e.>[]{lewin76,carson86,bartoldy02,rinaldi03}.
%...has been shown by \cite <e.g.,>[and others]{lewin76}.
%
% apacite uses < > for prenotes and [ ] for postnotes
% DO NOT use other cite commands (e.g., \citet, \citep, \citeyear, \nocite, \citealp, etc.).
%